\documentclass[useAMS,usenatbib,letterpaper,usegraphicx]{mn2e}

\usepackage{times,amssymb,hyperref,subfigure,epsfig,aas_macros}

\usepackage[totalwidth=480pt,totalheight=680pt,layoutvoffset=0.5cm]{geometry}

\newcommand{\alp}{$\alpha$ CrB}
\newcommand{\bet}{$\beta$ Tri}

\begin{document}

\title[Coplanar Circumbinary Debris Disks]{Coplanar circumbinary debris disks}

\author[G. M. Kennedy et. al.]{G. M. Kennedy\thanks{Email:
    \href{mailto:gkennedy@ast.cam.ac.uk}{gkennedy@ast.cam.ac.uk}}$^1$, M. C. Wyatt$^1$,
  B. Sibthorpe$^2$, N. M. Phillips$^3$,\newauthor
  B. C. Matthews$^{4,5}$, J. S. Greaves$^6$ \\
  $^1$ Institute of Astronomy, University of Cambridge, Madingley Road, Cambridge CB3 0HA, UK\\
  $^2$ UK Astronomy Technology Center, Royal Observatory, Blackford Hill, Edinburgh EH9
  3HJ, UK\\
  $^3$ Joint ALMA Observatory, Alonso de Córdova 3107, Vitacura - Santiago, Chile \\
  $^4$ National Research Council of Canada, 5071 West Saanich Road, Victoria, BC, Canada V9E 2E7\\
  $^5$ University of Victoria, Finnerty Road, Victoria, BC, V8W 3P6, Canada\\
  $^6$ School of Physics and Astronomy, University of St Andrews, North Haugh, St Andrews, Fife KY16 9SS, UK}

\maketitle

\begin{abstract}
  We present resolved \emph{Herschel} images of circumbinary debris disks in the \alp~(HD
  139006) and \bet~(HD13161) systems. By modelling their structure, we find that both
  disks are consistent with being aligned with the binary orbital planes. Though secular
  perturbations from the binary can bring the disk into alignment, in both cases the
  alignment time at the distances at which the disk is resolved is greater than the
  stellar age, so we conclude that the coplanarity was primordial. Neither disk can be
  modelled as a narrow ring, requiring extended radial distributions. To satisfy both the
  \emph{Herschel} and mid-IR images of the \alp~disk, we construct a model that extends
  from 1-300AU, whose radial profile is broadly consistent with a picture where
  planetesimal collisions are excited by secular perturbations from the binary. However,
  this model is also consistent with stirring by other mechanisms, such as the formation
  of Pluto-sized objects. The \bet~disk is modelled as a disk that extends from
  50-400AU. A model with depleted (rather than empty) inner regions also reproduces the
  observations and is consistent with binary and other stirring mechanisms. As part of
  the modelling process, we find that the \emph{Herschel} PACS beam varies by as much as
  10\% at 70$\mu$m and a few \% at 100$\mu$m. The 70$\mu$m variation can therefore hinder
  image interpretation, particularly for poorly resolved objects. The number of systems
  in which circumbinary debris disk orientations have been compared with the binary plane
  is now four. More systems are needed, but a picture in which disks around very close
  binaries (\alp, \bet, and HD 98800, with periods of a few weeks to a year) are aligned,
  and disks around wider binaries (99 Her, with a period of 50 years) are misaligned, may
  be emerging. This picture is qualitatively consistent with the expectation that the
  protoplanetary disks from which the debris emerged are more likely to be aligned if
  their binaries have shorter periods.
\end{abstract}


\section{Introduction}\label{s:intro}

The \emph{Herschel} Key Program DEBRIS (Dust Emission via a Bias free Reconnaissance in
the Infrared/Submillimeter) has observed a large sample of nearby stars to discover and
characterise extrasolar analogues to the Solar System's asteroid and Edgeworth-Kuiper
belts, collectively known as ``debris disks.'' The 3.5m \emph{Herschel} mirror diameter
provides 6-7'' resolution at 70-100$\mu$m \citep{2010A&A...518L...1P}, and as a
consequence our survey has resolved many disks around stars in the Solar neighbourhood
for the first time \citep[][Wyatt
et. al. 2012]{2010A&A...518L.135M,2011MNRAS.417.1715C,2012MNRAS.421.2264K}.\footnote{\emph{Herschel}
  is an ESA space observatory with science instruments provided by European-led Principal
  Investigator consortia and with important participation from NASA.}

Here we present resolved images of circumbinary disks in the \alp~and \bet~systems. These
systems are interesting because unlike most debris disk+binary systems, the binary orbits
are well characterised. The combination of a known orbit and resolved disk means we can
compare their relative inclination.

Our observations of a disk around the binary 99 Her \citep{2012MNRAS.421.2264K} were a
step toward building on the binary debris disk study of
\citet{2007ApJ...658.1289T}. Their \emph{Spitzer} study found that debris disks are
generally as common in binary systems as in single systems, but are less likely to reside
in systems with binary separations in the 3-30AU range \citep[see
also][]{2012ApJ...745..147R}. However, only some of their systems had detections at
multiple wavelengths to constrain the disk location and none were reported as resolved,
making the true dust location uncertain. Even in the case of dust detection at multiple
wavelengths, the true dust location is unknown because grains of different compositions
and sizes can have the same temperature at different stellocentric distances. In addition
to uncertainty in the dust location, only the projected sky separation of the binary (not
the binary semi-major axis) was generally known, adding further uncertainty.

Systems with resolved disks and well characterised binary orbits, such as 99 Her, \alp,
\bet, and HD 98800 \citep{2010ApJ...710..462A} remove these ambiguities. The dust
location and structure can be inferred within the context of the binary orbit, leading to
robust conclusions about whether the dust resides on stable orbits. One can also consider
whether perturbations from the binary play an important role in setting the dust
dynamics. For example, in the 99 Her system the disk position angle appears misaligned
with the binary line of nodes. This misalignment is best explained with particles on
polar orbits because these are stable for the stellar lifetime
\citep{2012MNRAS.421.2264K}. Another question is whether binary perturbations can
``stir'' the disk particles by increasing their inclinations and eccentricities,
eventually resulting in high enough relative velocities that collisions are
destructive. This process is analogous to the planet stirring model proposed by
\citet{2009MNRAS.399.1403M}, though may rely on vertical (inclination) stirring because
companions of comparable mass induce lower forced eccentricities than companions that are
much less massive than the star \citep{2012MNRAS.421.2264K,2004ApJ...609.1065M}.

An additional link can be made to star and planet formation. Young binary systems with
small to medium ($\lesssim$100AU) separations are thought to form coplanar with their
protoplanetary disks, because the disk torque aligns the binary orbit on timescales short
relative to the disk lifetime \citep[e.g.][]{2000MNRAS.317..773B}. Testing this
prediction is difficult because it is difficult to ascertain disk and binary orientations
at the distances of the nearest star forming regions
\citep[e.g.][]{2007prpl.conf..395M}. Because debris disks around older main-sequence
binaries should retain the same orientation as the protoplanetary disks from which they
emerged, these disks yield information on the outcome of star formation. The advantage is
that compared to star-forming regions, these systems are much closer to Earth and hence
larger on the sky and brighter, making disk and binary characterisation much easier.

Though planet formation may be hindered to some degree by high collision velocities
induced by binary perturbations \citep[e.g.][]{2004ApJ...609.1065M,2007MNRAS.380.1119S},
the discovery of several circumbinary planets shows that planets do indeed form around
binary stars \citep[e.g.][]{2011Sci...333.1602D,2012Natur.481..475W}. Few such systems
are known because binaries are generally avoided by radial velocity surveys
\citep[see][]{2005ApJ...626..431K,2009ApJ...704..513K}. However, the very existence of
circumbinary debris disks provides evidence that planet formation around binaries can
proceed to form at least 10-100km sized objects, which must exist to feed the observed
dust through collisions \citep[e.g.][]{2008ARA&A..46..339W,2010RAA....10..383K}. Further,
circumbinary disks are relatively common \citep{2007ApJ...658.1289T}, suggesting that
circumbinary planets may be no less unusual than their circumstellar equivalents.

This paper is laid out as follows. We first consider the stellar and orbital properties,
along with previous IR observations of the \alp~and \bet~systems (\S \ref{s:stars}). We
then show the \emph{Herschel} data (\S \ref{s:obs}) and simple models (\S \ref{s:mod}),
and then interpret these within the context of the expected dynamics (\S \ref{s:dyn}). We
discuss the results and conclude in \S\S \ref{s:disc} \& \ref{s:sum}.

\section{The binary systems}\label{s:stars}

\subsection{$\alpha$ Coronae Borealis}\label{ss:alp}

Among the ``alpha'' stars, \alp~(HD 139006, HIP 76267) is the only known eclipsing binary
\citep{1914ApJ....39..459S,1986AJ.....91.1428T}. Though some properties of this binary
have been known for over a century \citep[e.g.][]{1910PAllO...1...85J}, only recently has
the orientation of the orbit on the sky been constrained by the \emph{Hipparcos} mission
\citep[which also provides a system distance of
23pc,][]{1997ESASP1200.....P,2007A&A...474..653V}. The orbital elements are given in
Table \ref{tab:alpcrb}. The primary is an A0 dwarf, orbited by a G5 secondary
\citep{1986AJ.....91.1428T}. The age of the system is about 350Myr
\citep{2001ApJ...546..352S,2005ApJ...620.1010R,2012AJ....143..135V}.

This system was discovered to have an IR excess with IRAS \citep{1985PASP...97..885A},
and the disk has subsequently been detected with the \emph{Spitzer}
MIPS\footnote{Multi-band Imaging Photometer for \emph{Spitzer}
  \citep{2004ApJS..154....1W,2004ApJS..154...25R}} and IRS\footnote{Infra-Red
  Spectrograph \citep{2004ApJS..154...18H}} instruments
\citep{2006ApJS..166..351C,2005ApJ...620.1010R,2006ApJ...653..675S}. The proximity and
brightness mean this system has also been the target of high resolution mid-IR imaging
campaigns, with resolution of the disk at 11$\mu$m \citep{2010ApJ...723.1418M}. They may
have resolved the disk at 18$\mu$m, but ambiguity arising from artefacts of the observing
procedure meant that the results at this wavelength were unclear. The position angle of
the extension in the 11$\mu$m and first 18$\mu$m images is roughly 350$^\circ$, so
consistent with the (rather uncertain) binary line of nodes. The images also suggest that
the disk is edge on, so the inner part of the disk appears to be aligned with the binary
orbital plane.

\begin{table}
  \caption{\alp~system properties and 1$\sigma$ uncertainties
    \citep{1986AJ.....91.1428T,1997ESASP1200.....P}. The ascending node $\Omega$ is
    measured anti-clockwise from North. The longitude of pericenter is measured
    anti-clockwise from the ascending node. The semi-major axis is calculated from the
    period, masses, and system distance. The luminosities are derived from our
    SED fitting to the AB pair (\S \ref{ss:sed}), with the individual values found by
    scaling the \citet{1986AJ.....91.1428T} values of 74 and 0.8$L_\odot$. They derive
    a slightly  greater system distance (26pc), which accounts for
    their larger luminosities. The uncertainty in $L_{\rm A}$ is set by the SED
    fitting, and for $L_{\rm B}$ we use the 25\% uncertainty from
    \citet{1986AJ.....91.1428T} to account for uncertainty in the flux
    ratio.}\label{tab:alpcrb}
  \begin{tabular}{llll}
    \hline
    Parameter & Symbol (unit) & Value & Uncertainty \\
    \hline
    Semi-major axis & a (mas) & 8.66 & 0.09 \\
    Semi-major axis & a (AU) & 0.2 & 0.002 \\
    Eccentricity & e & 0.37 & 0.01 \\
    Inclination & i ($^\circ$) & 88.2 & 0.1 \\
    Ascending node & $\Omega$ ($^\circ$) & 330 & 20 \\
    Longitude of pericenter & $\omega$ ($^\circ$) & 311 & 2 \\
    Period & P (days) & 17.3599 & 0.0005 \\
    Distance & d (pc) & 23.0 & 0.15 \\
    $M_{\rm A}$ & $ (M_\odot)$ & 2.58 & 0.045 \\
    $M_{\rm B}$ & $ (M_\odot)$ & 0.92 & 0.025 \\
    $L_{\rm A}$ & $(L_\odot)$ & 59.4 & 1 \\
    $L_{\rm B}$ & $(L_\odot)$ & 0.6 & 0.15 \\
    \hline
\end{tabular}  
\end{table}

\subsection{$\beta$ Trianguli}\label{ss:bet}

The \bet~(HD 13161, HIP 10064) system was first recognised as a double lined
spectroscopic binary over a century ago \citep{1909ApJ....30..239M}. The orbital period
was then found to be 37 days, based on 12 radial velocity measurements. Further study has
refined the orbit \citep{1928ApJ....67..336S,1960PDAO...11..277E} and modern instruments
have since resolved the pair allowing a visual orbit to be derived
\citep{1995AJ....110..376H}. The orbit is now well characterised,\footnote{The Washington
  Double Star Catalogue \citep[WDS,][]{2001AJ....122.3466M} lists the orbit as
  ``definitive'' (grade 5)} with the orbital elements given in Table
\ref{tab:bettri}. The primary star has a A5IV spectral type \citep{2003AJ....126.2048G},
and as the class IV indicates has probably reached the end of the main-sequence. The
spectral type of the secondary is not known, but the mass suggests a mid F-type. There
remain uncertainties in this system; as discussed by \citet{1995AJ....110..376H}, the A5
spectral type is at odds with the $3.5M_\odot$ mass, and the mass ratio is inconsistent
with the luminosity ratio. The system is simply classed as 'Old' by
\citet{2007ApJ...658.1289T}. \citet{2012AJ....143..135V} derive an isochrone age of
730Myr, though this figure is systematically uncertain both due to a relatively small
luminosity ratio (so the age may be overestimated) and because \bet~lies outside the
range encompassed by some isochrone models. The exact age is not particularly important
for our purposes here, so we assume it is 730Myr.

Like \alp, this system was first discovered to have an IR excess with IRAS
\citep{1986PASP...98..685S}. It was observed with \emph{Spitzer} MIPS as part of a sample
of main-sequence binaries \citep{2007ApJ...658.1289T}, and also observed with IRS in the
two longer wavelength modules (14-38$\mu$m).

\begin{table}
  \caption{\bet~system properties and 1$\sigma$ uncertainties
    \citep{2000A&AS..145..215P}. The ascending node $\Omega$ is measured anti-clockwise
    from North. The longitude of pericenter is measured anti-clockwise from the ascending
    node. The total binary luminosity is derived from the SED in \S \ref{ss:sed}.}\label{tab:bettri}
  \begin{tabular}{llll}
    \hline
    Parameter & Symbol (unit) & Value & Uncertainty \\
    \hline
    Semi-major axis & a (mas) & 8.03 & 0.06 \\
    Semi-major axis & a (AU) & 0.312 & 0.002 \\
    Eccentricity & e & 0.433 & 0.004 \\
    Inclination & i ($^\circ$) & 130 & 0.52 \\
    Ascending node & $\Omega$ ($^\circ$) & 245.2 & 0.67 \\
    Longitude of pericenter & $\omega$ ($^\circ$) & 118.1 & 0.66 \\
    Period & P (days) & 31.3900 & 0.0002 \\
    Distance & d (pc) & 38.9 & 0.5 \\
    $M_{\rm A}$ & $ (M_\odot)$ & 3.5 & 0.25 \\
    $M_{\rm B}$ & $ (M_\odot)$ & 1.4 & 0.1 \\
    $L_{\rm AB}$ & $(L_\odot)$ & 74 & 1.4 \\
    \hline
\end{tabular}  
\end{table}

\section{Observations}\label{s:obs}

\subsection{\emph{Herschel}}

\begin{table}
  \caption{\emph{Herschel} observations of \alp~and \bet.}\label{tab:obs}
  \begin{tabular}{lllll}
    \hline
    Target & ObsId & Date & Instrument & Duration (s) \\
    \hline
    \alp & 1342213794 & 7 Feb 2011 & PACS 100/160 & 445 \\
    \alp & 1342213795 & 7 Feb 2011 & PACS 100/160 & 445 \\
    \bet & 1342223650 & 4 July 2011 & PACS 100/160 & 445 \\
    \bet & 1342223651 & 4 July 2011 & PACS 100/160 & 445 \\
    \alp & 1342223846 & 9 July 2011 & PACS 70/160 & 445 \\
    \alp & 1342223847 & 9 July 2011 & PACS 70/160 & 445 \\
    \bet & 1342237390 & 12 Jan 2012 & PACS 70/160 & 445 \\
    \bet & 1342237391 & 12 Jan 2012 & PACS 70/160 & 445 \\
    \bet & 1342237504 & 14 Jan 2012 & SPIRE 250/350/500 & 721 \\
    \hline
  \end{tabular}  
\end{table}

\emph{Herschel} Photodetector and Array Camera \& Spectrometer
\citep[PACS,][]{2010A&A...518L...2P} data were taken for both \alp~and \bet~at 100 and
160$\mu$m during routine DEBRIS observations. Subsequently, a Spectral and Photometric
Imaging Receiver \citep[SPIRE,][]{2010A&A...518L...3G} observation of \bet~was triggered
by the large PACS excess indicating a likely sub-mm detection. For both targets we also
obtained 70$\mu$m PACS images to better resolve the disks. Because every PACS observation
includes the 160$\mu$m band, we have two images at this wavelength for each target. All
observations were taken in the standard scan-map modes for our survey; mini scan-maps for
PACS data and small maps for SPIRE. Data were reduced using a near-standard pipeline with
the Herschel Interactive Processing Environment \citep[HIPE Version
7.0,][]{2010ASPC..434..139O}. We decrease the noise slightly by including some
``turn-around'' data taken as the telescope is accelerating and decelerating at the start
and end of each scan leg.

\begin{figure*}
  \begin{center}
    \hspace{-0.5cm} \includegraphics[width=0.5\textwidth]{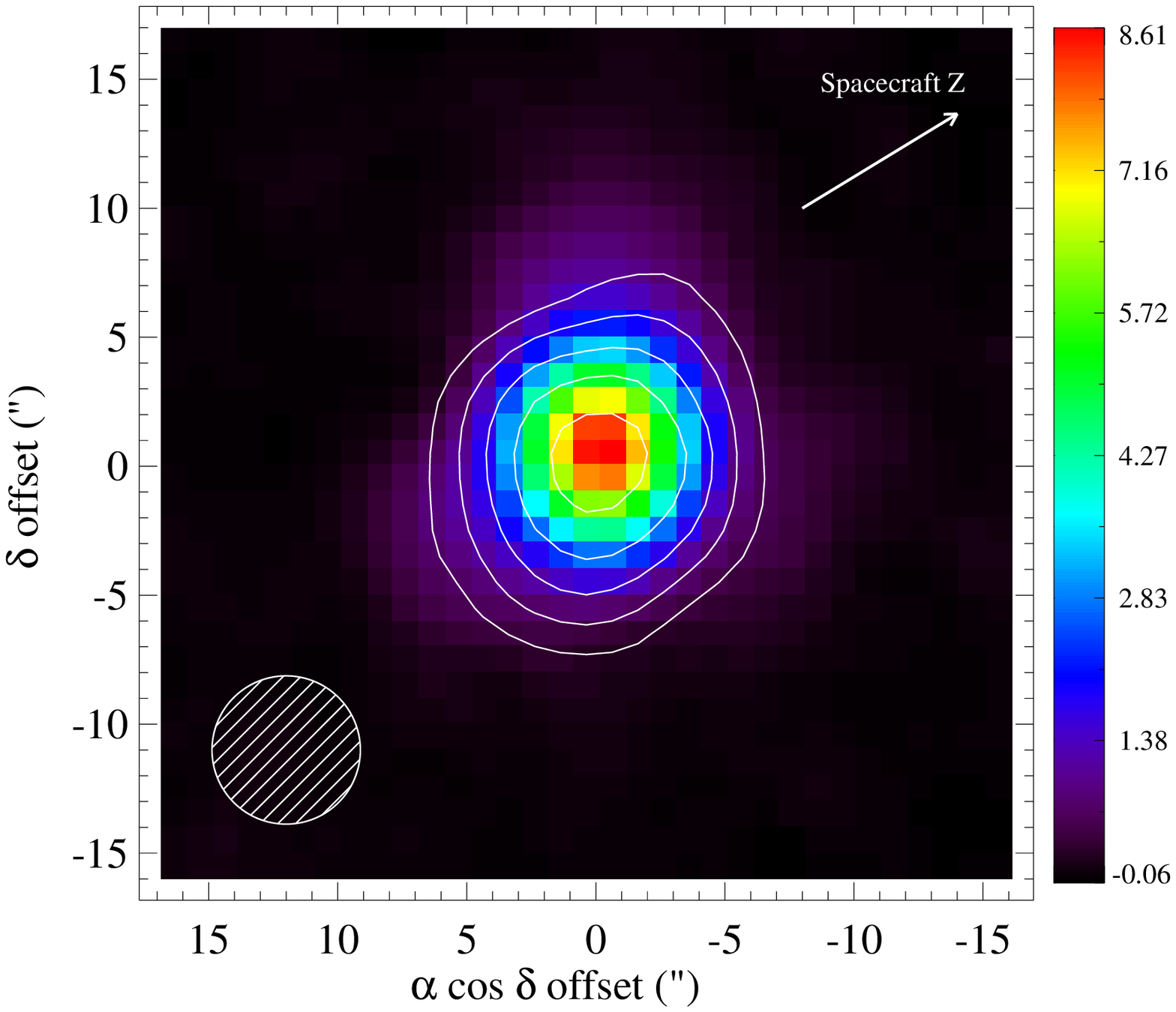}
    \includegraphics[width=0.5\textwidth]{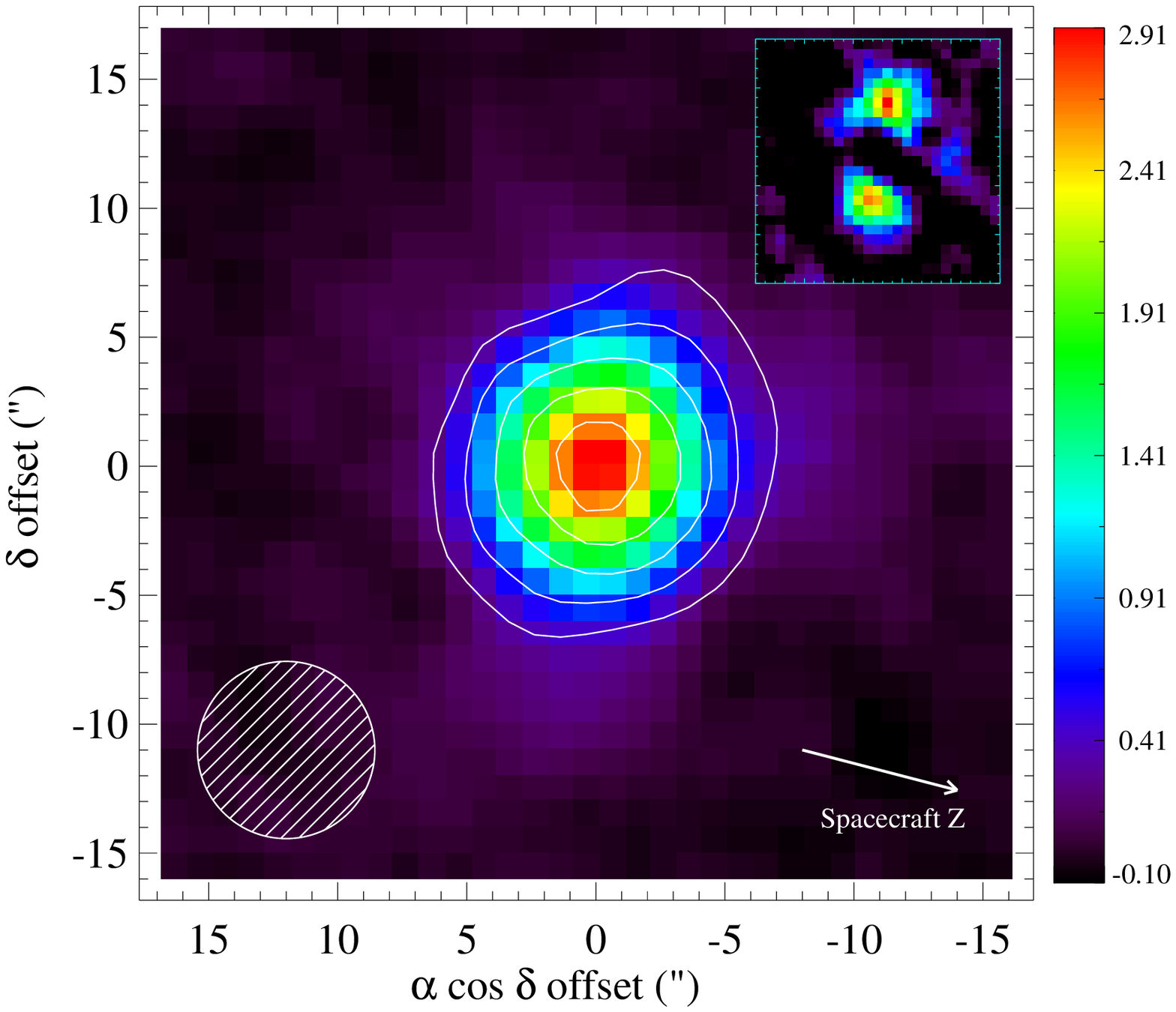}
    \caption{Resolved PACS images of \alp~at 70 (left) and 100$\mu$m (right). North is up
      and East is left, and the arrow shows the spacecraft ``Z'' direction. The colour
      scale is in mJy/square arcsecond. The overlaid contours show the corresponding
      160$\mu$m images in 5 linear steps from 5-13$\sigma$. Similar contours correspond
      to 30-200$\sigma$ for the 70$\mu$m image, and 10-75$\sigma$ for the 100$\mu$m
      image. The hatched circles show the average PACS beam FWHM of 5\farcs75 (70$\mu$m)
      and 6\farcs87 (100$\mu$m, see text). The inset (25'' square) shows residuals of the
      100$\mu$m \alp~image after PSF fitting, the lobes show that the disk is clearly
      resolved.}\label{fig:imgalp}
  \end{center}
\end{figure*}

Figure \ref{fig:imgalp} shows the \emph{Herschel} PACS images of \alp. The disk is not
obviously resolved, so the inset shows the residuals from a point-spread function (PSF)
fit to the 100$\mu$m data (an observation of \emph{Herschel} calibrator $\gamma$ Dra
generated using the same data reduction method was used as a PSF, see below). The
residual emission in the inset is symmetric about the stellar position along a position
angle of about 350$^\circ$, showing that the disk is most likely resolved with a near
edge-on geometry. The emission extends to around 10'' ($\sim$230AU) from either side of
the stellar position, suggesting a disk diameter of several hundred AU. The disk does not
appear to be resolved along the minor axis, so is consistent with the disk being edge on
and aligned with the (eclipsing) binary plane. A PSF fit to the 70$\mu$m image shows
similar results, while the 160$\mu$m image is unresolved.

Because the \alp~disk is not well resolved, it is possible that this extension is
affected by variation in the PACS PSF (or beam). Further, a key difference between our
method and that used for beam characterisation by the PACS team is that because they are
much fainter than the calibration stars our data cannot be re-centered on a
frame-by-frame basis, resulting in a slightly larger beam size. Thus, comparison of the
Gaussian size in our images with those reported by the PACS team could lead to the
conclusion that the disk is slightly more extended than it really is. To characterise the
beam variation we have obtained two observations each of the five PACS calibration stars
at both 70 and 100$\mu$m. These observations were nearly all taken using the same mini
scan-map mode used by DEBRIS. These data are reduced in the same way as all DEBRIS data,
so allow a realistic comparison. The expectation is that each calibration PSF will be
slightly different, with the differences across all ten being representative of the
uncertainty in the PSF specific to a given science observation.

\begin{figure*}
  \begin{center}
    \hspace{0cm} \includegraphics[width=1\textwidth]{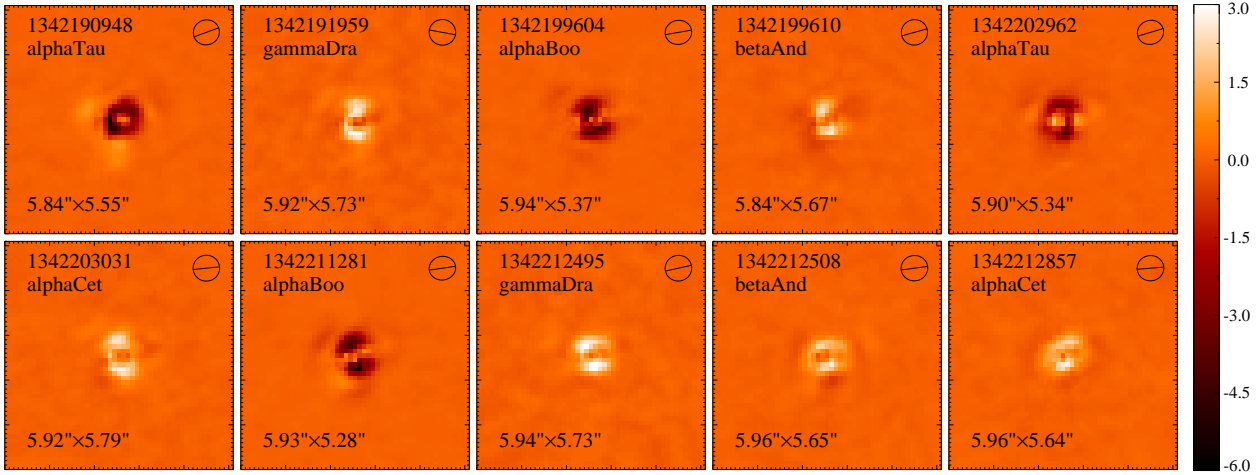}
    \caption{PACS 70$\mu$m beam comparison. Each panel shows the beam after the average
      beam has been subtracted, with labels noting ObsID, star name, and Gaussian major
      and minor FWHM. The ellipse shows the FWHM, with the line along the major axis. The
      colour scale shows the fractional difference (in \%) relative to the peak. In these
      images the spacecraft ``Z'' direction is up and ``Y'' is to the right, and the two
      scan directions that make a mini scan map are $\pm20^\circ$ from
      horizontal.}\label{fig:psf70}
  \end{center}
\end{figure*}

The 70$\mu$m calibration observations are shown in Figure \ref{fig:psf70}, where each
image shows the residuals when the average of all ten observations is subtracted (after
the peak is scaled to the same as the observation). All images have been rotated so that
spacecraft ``Z'' is up (spacecraft ``Y'' is to the right). Each panel shows the
observation number (ObsID), star name, and Gaussian major and minor FWHM. The ellipse
shows the Gaussian FWHM and the line shows the major axis orientation. Because the
expectation is that each observation is independent, the panels are ordered
chronologically (i.e. by ObsID). It is clear that the beam width varies at about the 10\%
level along the minor axis, between 5\farcs28 for $\alpha$ Boo and 5\farcs79 for $\alpha$
Cet. The variation along the major axis is much smaller at about 2\%. The beam variation
appears to be systematically different for two stars; both the $\alpha$ Boo and $\alpha$
Tau images show similarly small minor FWHM despite being observed about six months
apart. This variation cannot be attributed to circumstellar material because none of the
calibration stars have IR excesses. It may be difficult to confirm this possible
systematic effect, for example as a function of Solar elongation or declination, as there
are only five PACS calibration stars.

The major and minor FWHM of the average 70$\mu$m PACS beam are 5\farcs91 and 5\farcs59,
so similar to, but slightly larger than, the values of 5\farcs76 and 5\farcs46 found by
the PACS team (for a scan speed of 20''/sec).\footnote{See PACS Observer's Manual} Based
on the variation seen in Figure \ref{fig:psf70}, the uncertainties on these values are
about 0\farcs05 in the major axis (approximately along the spacecraft ``Y'' direction)
and 0\farcs2 in the minor axis (approximately along the spacecraft ``Z'' direction).

\begin{figure*}
  \begin{center}
    \hspace{-0.0cm} \includegraphics[width=1\textwidth]{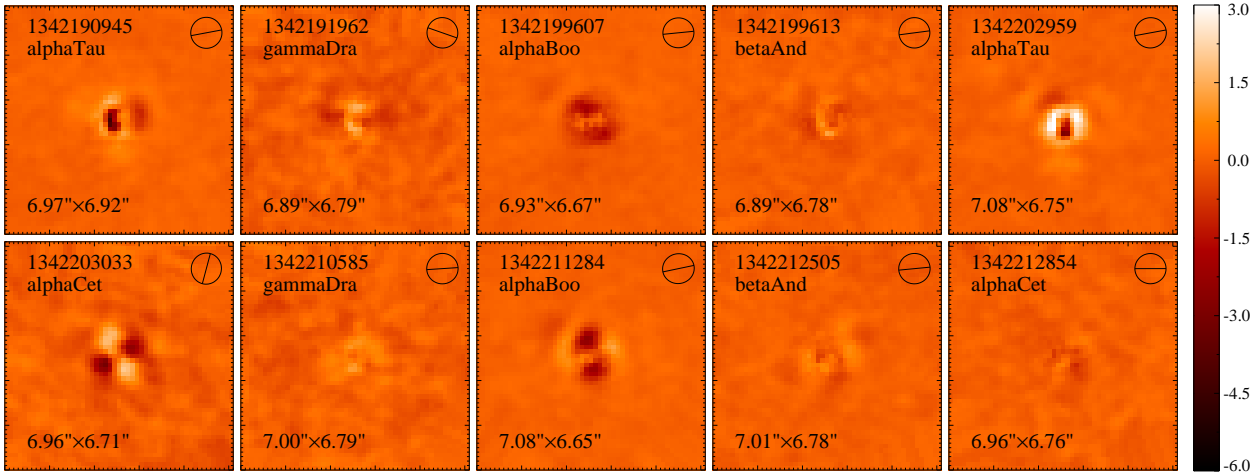}
    \caption{PACS 100$\mu$m beam comparison. The layout is as described in Figure
      \ref{fig:psf70}}\label{fig:psf100}
  \end{center}
\end{figure*}

A similar analysis for the 100$\mu$m beam is shown in Figure \ref{fig:psf100}. The panels
are again in chronological order, so the order of the stars is not the same as at
70$\mu$m. The variation is smaller at this wavelength (2-4\%), but some systematic
difference for $\alpha$ Tau and $\alpha$ Boo is still seen. The average major and minor
beam FWHM are 6\farcs95 and 6\farcs78 (compared to 6\farcs89 and 6\farcs69 found by the
PACS team), with uncertainties of about 0\farcs1 in both axes. The lower left panel
($\alpha$ Cet, ObsID:1342203033) shows rather different residuals compared to all others,
arising because the observation comprises only one scan direction. This difference shows
that aside from variation, the beam is also a function of the observing strategy, and
that the strategy should be the same when using calibration stars to analyse science data
(e.g. for PSF fitting and image modelling).

Having quantified the beam characteristics specific to our data, we now compare them to
the \alp~observations. A Gaussian fit to the star-subtracted 70$\mu$m image finds a
position angle of $348 \pm 3^\circ$, which is consistent with the ascending node of $330
\pm 20^\circ$ for the binary. The major and minor FWHM are $7\farcs3 \pm 0\farcs1$ and
$6\farcs1 \pm 0\farcs1$. Figure \ref{fig:psf70} shows that a major axis of the beam is
typically perpendicular to the spacecraft Z axis, so the disk position angle is between
the major and minor axes, where the beam has a FWHM of about
$5.75\pm0\farcs1$. Therefore, the disk is clearly resolved along the direction of the PA,
but at only about 2$\sigma$ significance in the perpendicular direction. Simple
deconvolution suggests a disk diameter of about 100AU. This size is much smaller than
suggested by the residuals in the inset in Figure \ref{fig:imgalp}, indicating that the
bulk of the emission is poorly resolved. This structure is perhaps consistent with the
mid-IR imaging if \alp~hosts warm and cold components that both contribute to the
\emph{Herschel} fluxes but only the cold component is resolved with
\emph{Herschel}. Though the minor Gaussian FWHM is slightly larger than expected,
variation in the 70$\mu$m beam means that the disk is consistent with being edge-on, and
therefore aligned with the (eclipsing) binary orbital plane. This analysis also shows
that the outer component resolved by \emph{Herschel} is consistent with being aligned
with the inner component resolved in the mid-IR \citep{2010ApJ...723.1418M}. At 100$\mu$m
the major and minor FWHM are $8\farcs2 \pm 0\farcs16$ and $6\farcs9 \pm 0\farcs13$,
yielding a size of about 100AU and a position angle (PA) of $347 \pm 4^\circ$. Therefore,
at this wavelength the disk is resolved along the disk PA, but not in the perpendicular
direction, so is again consistent with being edge-on, and aligned with the binary
plane. We return to the issue of beam variation when creating resolved models of the
\alp~disk in \S \ref{ss:img}.

\begin{figure*}
  \begin{center}
    \hspace{-0.5cm} \includegraphics[width=0.5\textwidth]{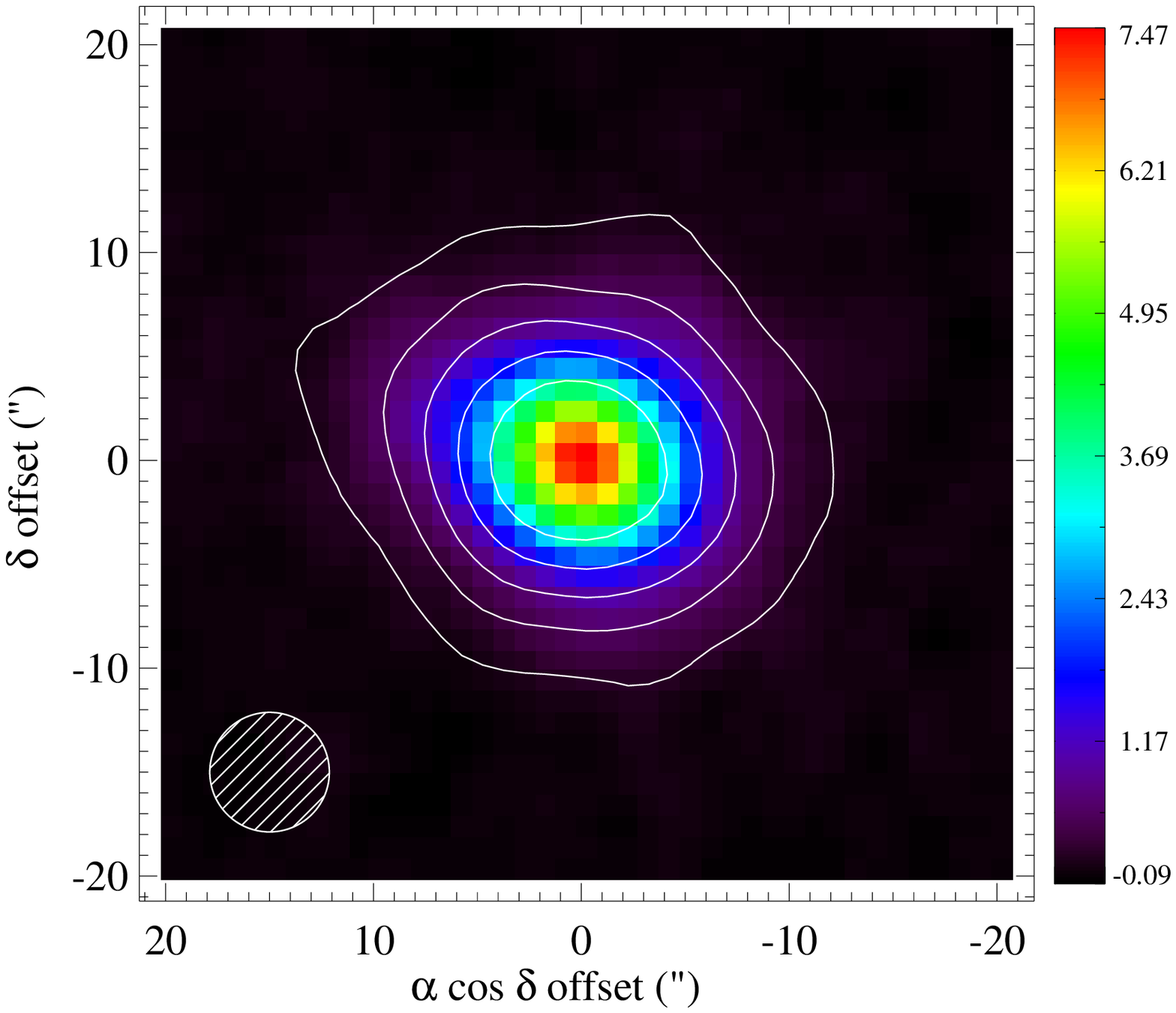}
    \includegraphics[width=0.5\textwidth]{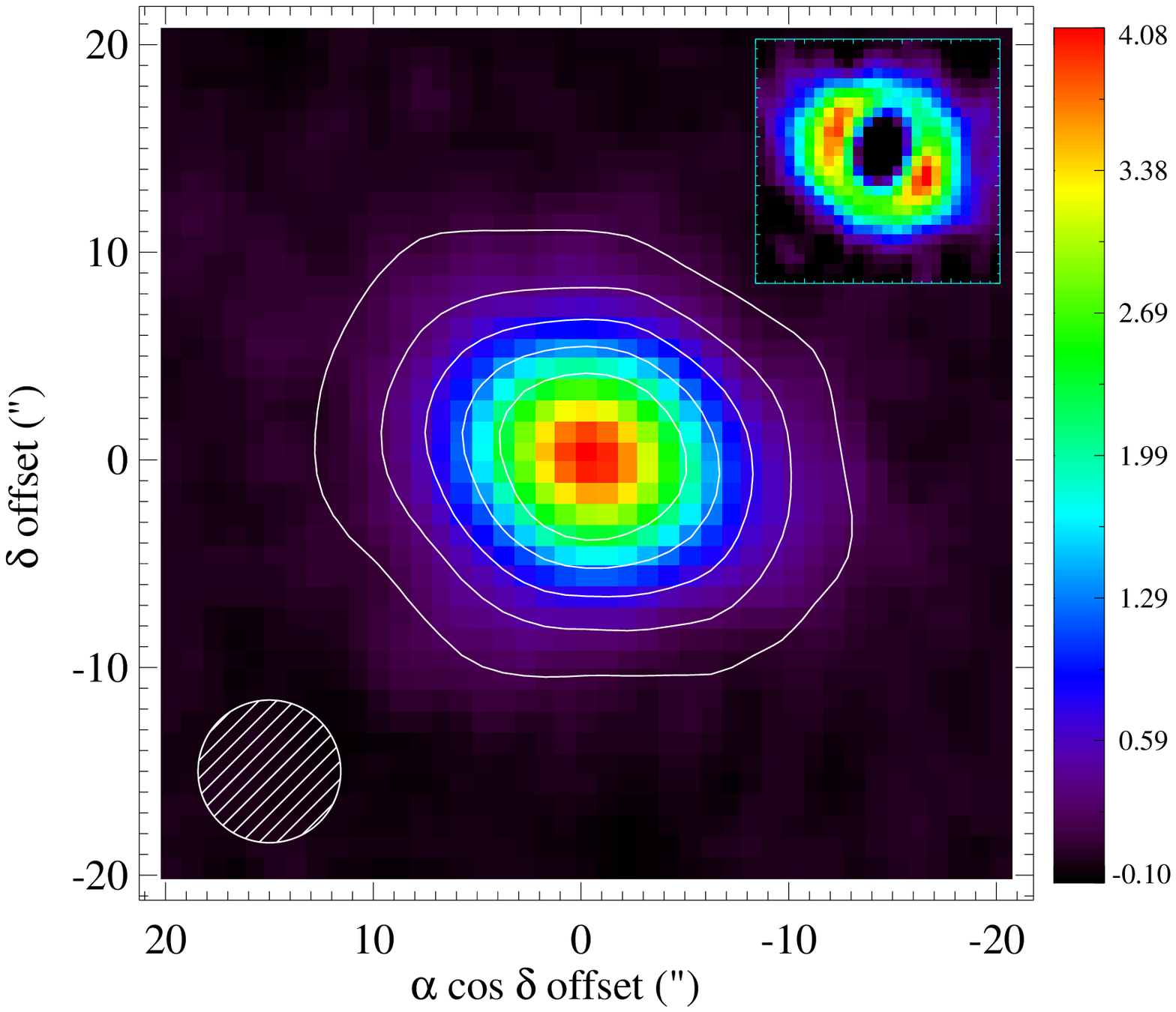}
    \caption{Resolved PACS images of \bet~at 70 (left) and 100$\mu$m (right). North is up
      and East is left. The colour scale is in mJy/square arcsecond. The overlaid
      contours show the corresponding 160$\mu$m images in 5 linear steps from
      5-30$\sigma$. Similar contours correspond to 5-75$\sigma$ for the 70$\mu$m image,
      and 5-70$\sigma$ for the 100$\mu$m image. The hatched circles show the average PACS
      beam FWHM of 5\farcs75 and 6\farcs87 (see text). The inset (25'' square) shows
      residuals of the 100$\mu$m \bet~image after PSF fitting, the elliptic residual
      emission shows that the disk is clearly resolved.}\label{fig:imgbet}
  \end{center}
\end{figure*}

The \bet~images are shown in Figure \ref{fig:imgbet}. Compared to the beam size, the disk
around \bet~is clearly resolved at 70 and 100$\mu$m. The disk is also resolved at
160$\mu$m, but not at the SPIRE wavelengths of 250-500$\mu$m. A Gaussian fit to the
star-subtracted image at 70$\mu$m finds major and minor FWHM of $9\farcs1 \pm 0\farcs13$
and $7\farcs5 \pm 0\farcs1$ at a PA of $66 \pm 3^\circ$. Simple Gaussian deconvolution
suggest a disk size of about 270AU, and an inclination of about $46 \pm 3^\circ$. At
100$\mu$m the FWHM are $10\farcs4 \pm 0\farcs13$ and $8\farcs6 \pm 0\farcs1$ at a PA of
$68 \pm 3^\circ$, suggesting a size of about 300AU and inclination of $48 \pm
3^\circ$. At 160$\mu$m the FWHM are $14\farcs4 \pm 0\farcs12$ and $12\farcs3 \pm
0\farcs1$ at a PA of $68 \pm 2^\circ$ (the 160$\mu$m beam is elongated and about
10\farcs7 by 12\farcs1). The 160$\mu$m image is therefore poorly resolved, but suggests
an approximate size of 300-370AU. Taken together, these angles mean that in this system
the disk is again consistent with being aligned with the binary plane. The increasing
disk size with wavelength suggests that the disk may be extended.

The flux density in the PACS images for both \alp~and \bet~is measured using apertures
and the SPIRE fluxes with PSF fitting. The measurements are given in Table
\ref{tab:phot}. The uncertainties include both statistical and systematic uncertainties
due to repeatability and calibration \citep[see][for further comments on
calibration]{2012MNRAS.421.2264K}.

\begin{table}
  \caption{\emph{Herschel}  photometry and 1$\sigma$ uncertainties of \alp~and \bet. The
    160$\mu$m measurements from each observation are given separately.}\label{tab:phot}
  \begin{tabular}{llrrl}
    \hline
    Target & Band & Flux (mJy) & Uncertainty (mJy) & Method \\
    \hline
    \alp & PACS70 & 515 & 26 & 15'' aperture \\
    \alp & PACS160 & 69 & 5 & PSF fit\\
    \alp & PACS100 & 235 & 12 & 15'' aperture \\
    \alp & PACS160 & 66 & 5 & PSF fit\\
    \bet & PACS70 & 641 & 32 & 20'' aperture \\
    \bet & PACS160 & 264 & 17 & 25'' aperture \\
    \bet & PACS100 & 481 & 23 & 20'' aperture \\
    \bet & PACS160 & 264 & 16 & 25'' aperture \\
    \bet & SPIRE250 & 87 & 7 & PSF fit \\
    \bet & SPIRE350 & 35 & 5 & PSF fit \\
    \bet & SPIRE500 & 5 & 5 & PSF fit \\
    \hline
  \end{tabular}  
\end{table}

\subsection{\emph{Spitzer}}

As noted above, both systems were observed by \emph{Spitzer} with the MIPS and IRS
instruments. The MIPS photometry is taken from \citet{PhillipsThesis}, and the IRS
spectra retrieved from the Cornell Atlas of Spitzer/Infrared Spectrograph Sources
\citep[CASSIS,][]{2011ApJS..196....8L}. The PACS and MIPS 70$\mu$m measurements are
consistent ($460 \pm 47$mJy for \alp~and $677 \pm 62$mJy for \bet). Though we instead use
the PACS images, \bet~is clearly resolved at 70$\mu$m with \emph{Spitzer}.

\section{Models}\label{s:mod}

\subsection{SEDs}\label{ss:sed}

\begin{figure*}
  \begin{center}
    \hspace{-0.5cm} \includegraphics[width=0.5\textwidth]{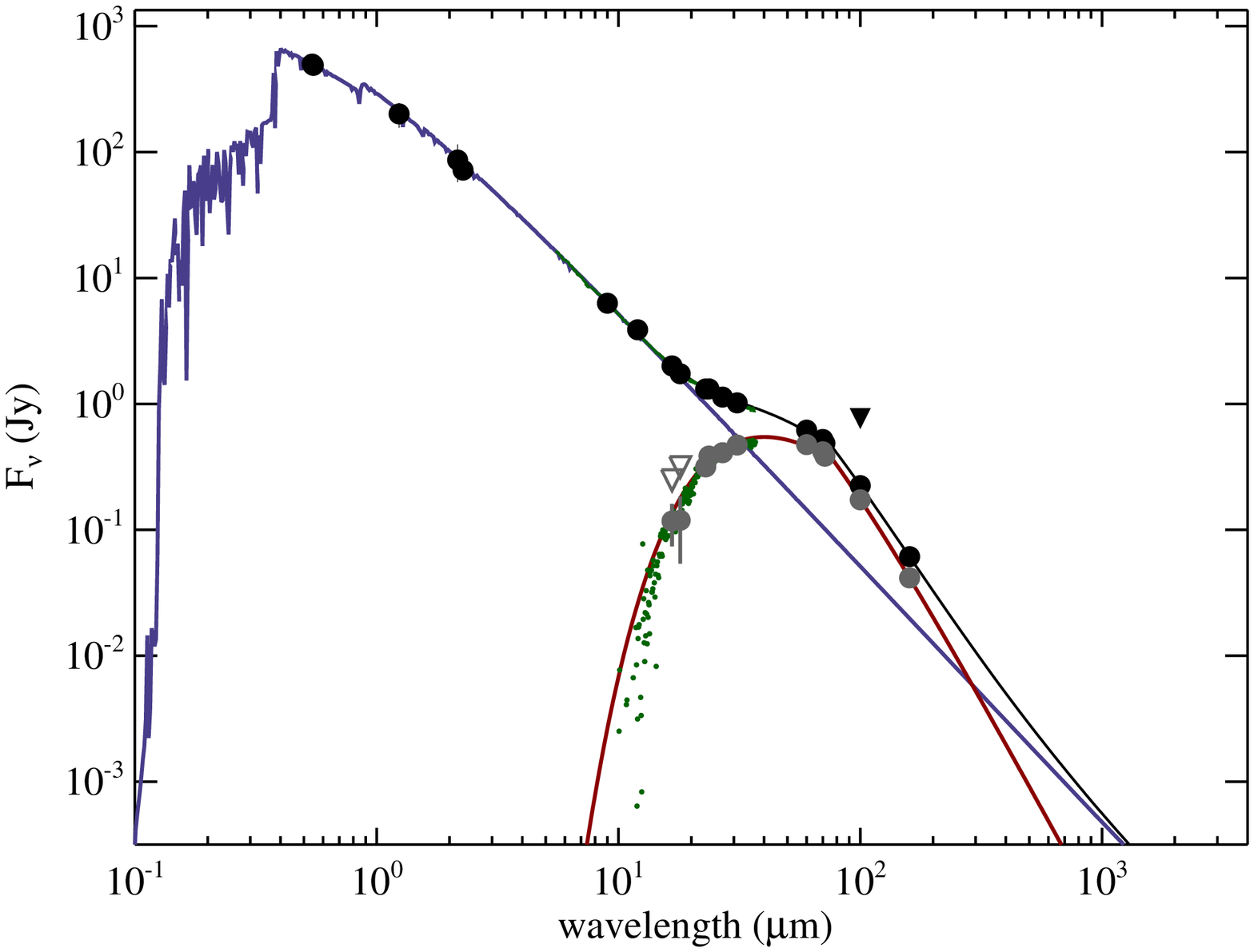}
    \hspace{0.1cm} \includegraphics[width=0.5\textwidth]{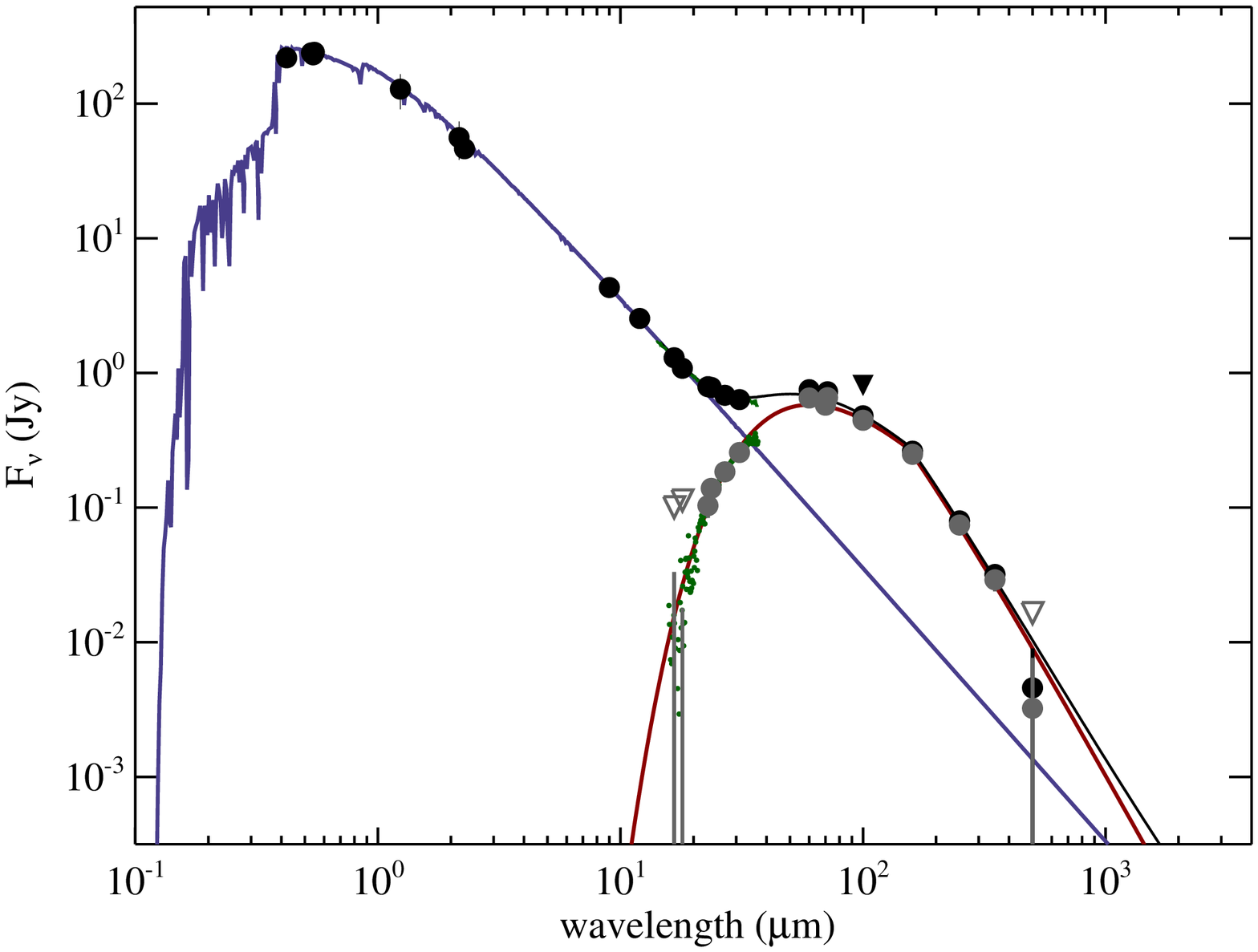}
    \caption{SEDs for \alp~(left) and \bet~(right). Photometry is shown as black dots or
      black triangles for upper limits. Disk (i.e. photosphere-subtracted) fluxes and
      upper limits are shown as grey dots, open triangles, and small green dots for
      IRS. The stellar spectrum is shown as a blue line and the blackbody disk model as a
      red line, with the total shown as a black line.}\label{fig:sed}
  \end{center}
\end{figure*}

We first model the stellar and disk photometry to derive approximate disk temperatures
and fractional luminosities. This photometry is collected from numerous catalogues
\citep{1978A&AS...34..477M,1993yCat.2156....0M,1997yCat.2215....0H,1997ESASP1200.....P,2000A&A...355L..27H,2003tmc..book.....C,2006yCat.2168....0M,2010A&A...514A...1I,PhillipsThesis,2011ApJS..196....8L}. Photometry
and colours at wavelengths up to 9$\mu$m are used to model the stellar photosphere, using
the PHOENIX Gaia grid \citep{2005ESASP.576..565B}. The best fit model is found by least
squares minimisation. At wavelengths longer than 9$\mu$m we model the excess emission
above the photosphere with a simple modified blackbody function; at wavelengths beyond
$\lambda_0$ the blackbody is multiplied by $\left(\lambda/\lambda_0\right)^{-\beta}$.

The spectral energy distributions (SEDs) are shown in Figure \ref{fig:sed}. The best
fitting stellar model for \alp~has $T_{\rm eff}=9280 \pm 100$K, $R_\star=3.06 \pm 0.03
R_\odot$, and $L_\star=60 \pm 1 L_\odot$, and for \bet~has $T_{\rm eff}=8000 \pm 100$K,
$R_\star=4.6 \pm 0.05 R_\odot$, and $L_\star=74 \pm 2 L_\odot$. Because no photometry
resolves either binary, these parameters simply describe what the total stellar emission
spectra look like, and are not physical. The \alp~disk model has temperature $T_{\rm
  disk}=124$K, fractional luminosity $f=1.7 \times 10^{-5}$, corresponding to a total
grain surface area $\sigma_{\rm tot}=0.34$AU$^2$ for blackbody grains, and $\lambda_0=75
\mu$m and $\beta=1.6$. The \bet~model has temperature $T_{\rm disk}=84$K, $f=3.0 \times
10^{-5}$, total grain surface area $\sigma_{\rm tot}=3.3$AU$^2$ for blackbody grains, and
$\lambda_0=152 \mu$m and $\beta=1.25$.

Despite the suggestion of warm dust by mid-IR imaging \citep{2010ApJ...723.1418M}, the
\alp~disk SED, most notably the IRS spectrum, does not require dust at multiple
temperatures. The inferred disk radius assuming blackbody properties for the single
component fit to the SED is 40AU, well beyond the few AU scale of the mid-IR
emission. The SED does not strongly preclude the existence of warm dust however, because
it is limited by the $\sim$2\% level that is achievable with typical photometric
calibration, and a similar uncertainty for the IRS spectrum. Therefore, at 10-20$\mu$m an
excess of $\lesssim$100mJy could be present yet not confidently detected. How this limit
converts to a limit on the fractional luminosity depends on the dust temperature
\citep[e.g.][]{2008ARA&A..46..339W}. In the case of 300K dust the limit on the fractional
luminosity is about $10^{-5}$, so only about a factor of two lower than derived from the
124K disk SED above.

For \bet~the blackbody radius is larger at around 95AU. This size is somewhat smaller
than the 150AU derived from the Gaussian fitting, but as noted above the disk could be
extended. Aside from being marginally resolved with MIPS, there are no constraints on the
structure from other observations.

\subsection{Images}\label{ss:img}

The main goal of modelling the resolved images is to derive the orientation of the disk
in space, for comparison with the orientation of the binary orbital plane. Though
alignment is already suggested by Gaussian fitting in both cases, the PACS beam is not
azimuthally symmetric so resolved modelling provides a worthwhile check. To achieve this
goal, we must find a spatial structure that reproduces the images satisfactorily. The
models are generated using the method described in \citet{2012MNRAS.424.1206W} \citep[see
also][]{2012MNRAS.421.2264K}. Basically, a high resolution model of the structure is
generated and multiplied by the grain emission properties at each wavelength. The
structure is the spatial distribution of dust, implemented as the cross-sectional area as
a function of radial distance from the star. The distribution is disk-like with a small
range of particle inclinations and the dust's face-on optical depth ($\tau$) is
parameterised as a function of radius by one or more power laws. The grain emission
properties are simply described with a modified blackbody whose temperature decreases
with radial stellocentric distance as a power-law. The high resolution models are then
created with some spatial orientation and convolved with the instrument beam for
comparison with the observed images. The large number of model parameters, as well as
different possible configurations (e.g. extended disks vs. multiple rings) means that we
do not undertake a grid search of possible parameter spaces. The final models are found
by a combination of by-eye fitting and least-squares minimisation.

Our general approach to modelling is to use the simplest model possible to explain the
data in hand, increasing the model complexity as required. Here, we find that the
simplest model, a narrow ring, is not sufficient to reproduce the PACS images for either
system. As a second step we use extended dust distributions, though two discrete narrow
rings is also a possibility (e.g. Wyatt et al. 2012), as is the inclusion of an
unresolved component. Because we do not use a grid approach, estimating uncertainties on
model parameters, many of which are highly degenerate, is made more difficult, However,
we find that multiple different configurations can reasonably reproduce the data, which
gives an idea of how well the data constrain the disk structure given the relatively poor
resolution compared to the disk sizes. Similarly, we find no requirement for the
temperature distribution to deviate from a simple blackbody relation ($T = 278.3
L_\star^{1/4} r^{-0.5}$K, where $L_\star$ is the binary luminosity in Solar units, and
$r$ is the disk radius in AU). We do not claim this is the true temperature distribution,
given that grain temperatures are expected to be somewhat different (hotter) than the
blackbody relation suggests, but that there is insufficient information to quantify this
difference.

\subsubsection{\alp}

\begin{figure*}
  \begin{center}
    \hspace{-0.5cm} \includegraphics[width=0.33\textwidth]{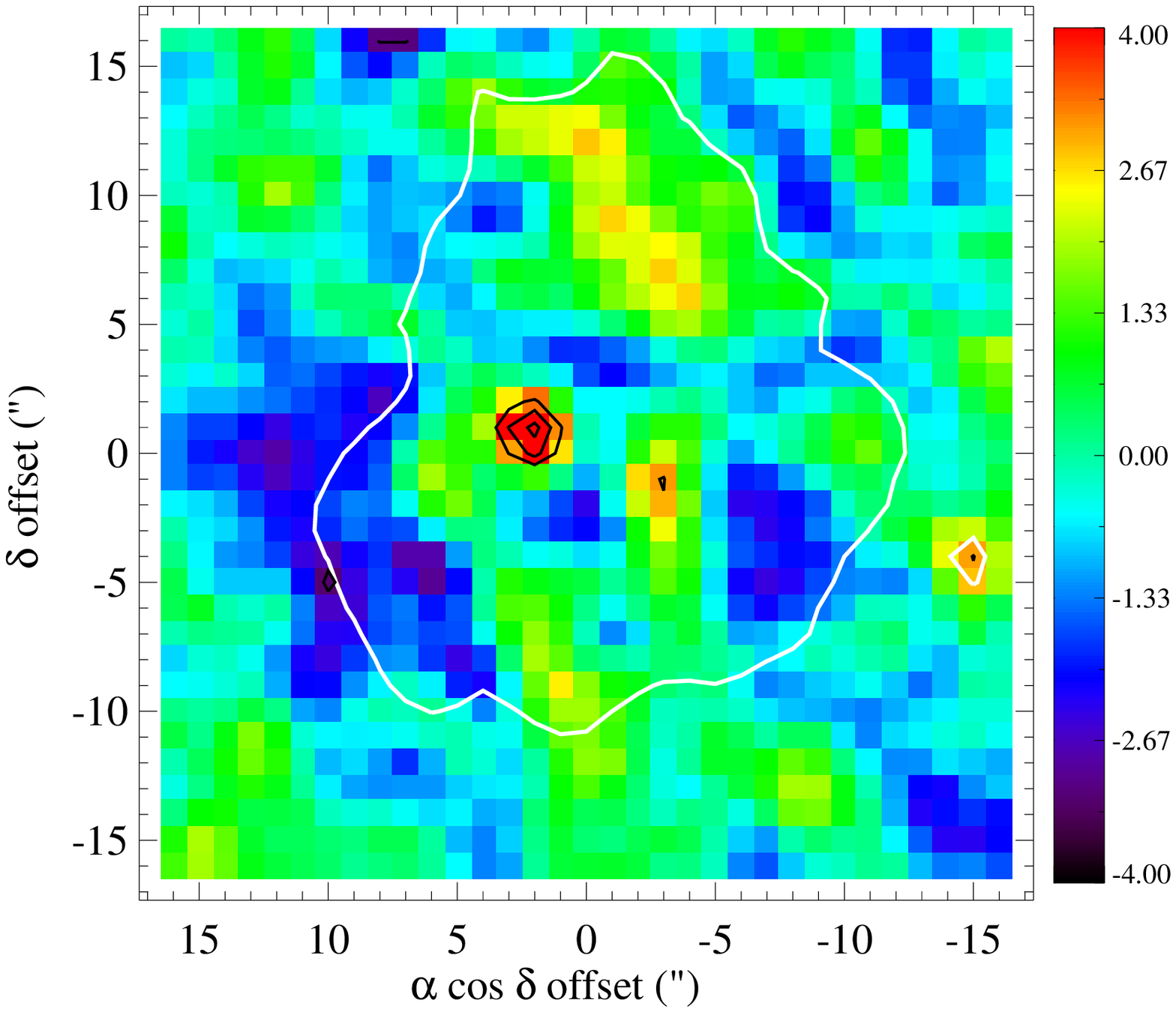}
    \hspace{0.1cm} \includegraphics[width=0.33\textwidth]{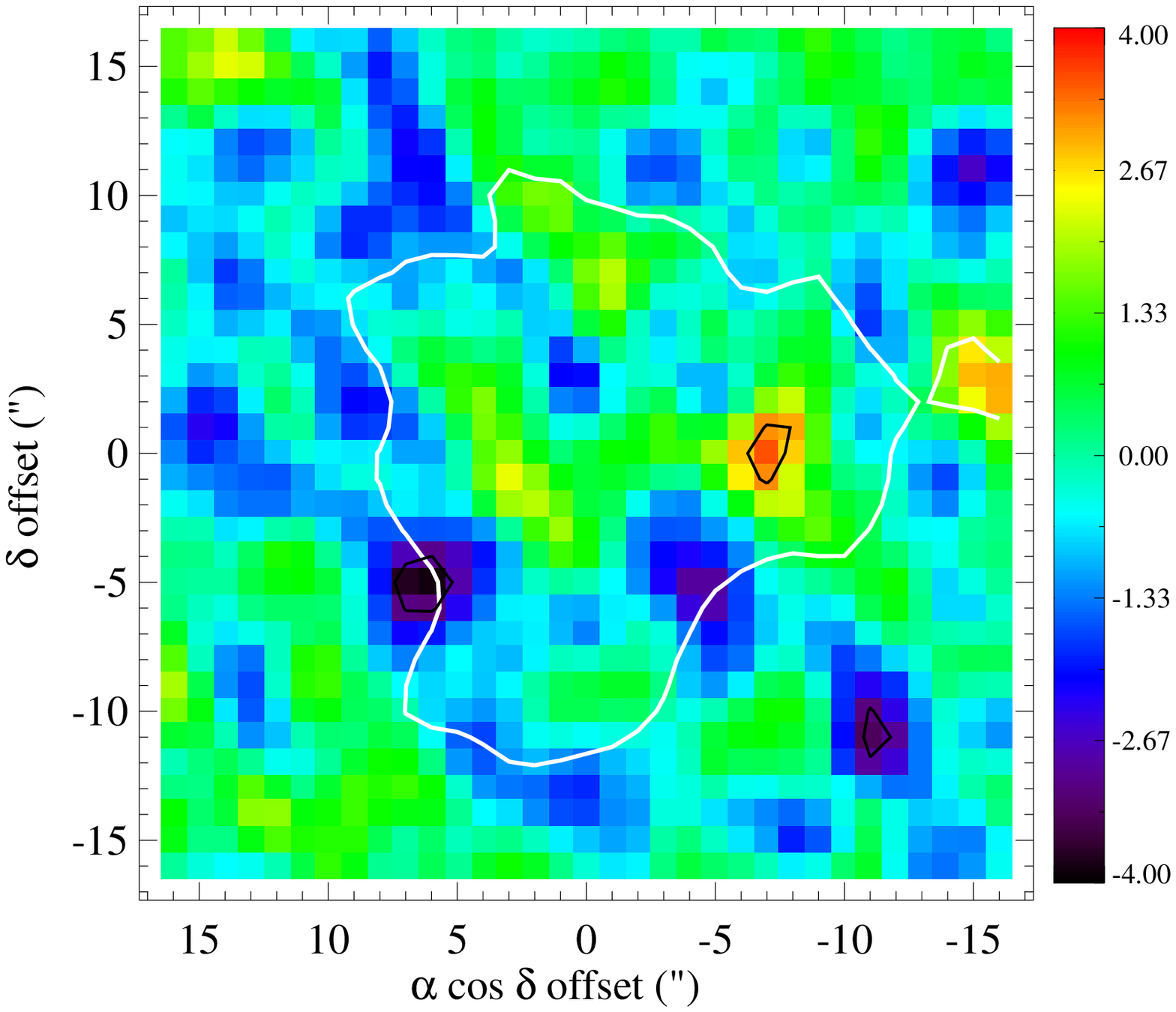}
    \hspace{0.1cm} \includegraphics[width=0.33\textwidth]{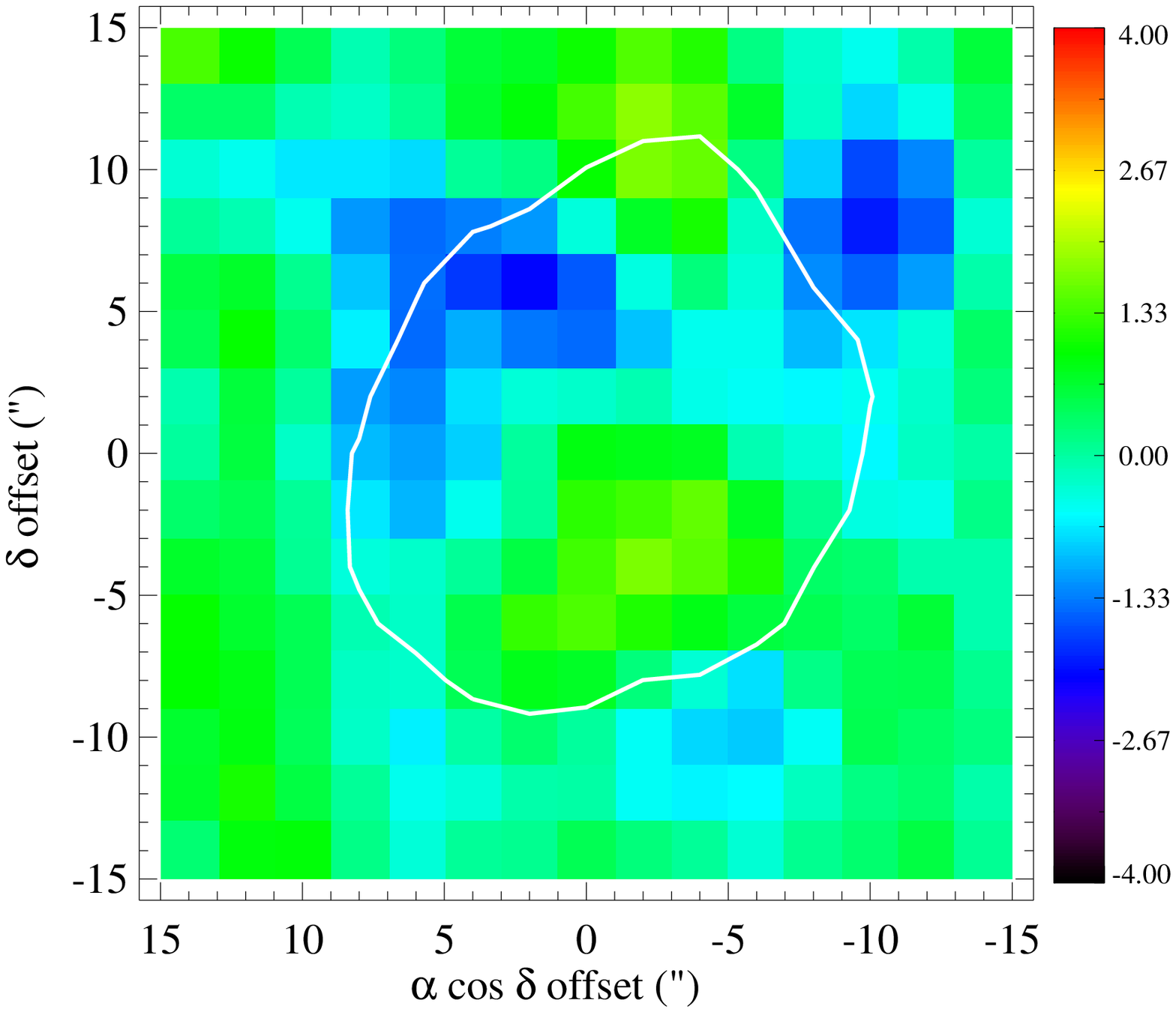}
    \caption{Residuals after the resolved model of \alp~is subtracted from the
      observations at 70$\mu$m (left) 100$\mu$m (middle) and 160$\mu$m (right). The
      colour scale is in units of the pixel to pixel uncertainty, and contours are shown
      at $\pm$3, 4 and 5$\sigma$. The white contour shows the 3$\sigma$ contour from the
      observed images.}\label{fig:alpres}
  \end{center}
\end{figure*}

We first model \alp, for which about half to two thirds of the \emph{Herschel} emission
is unresolved, with the remainder more distant from the stellar position and resolved
(see Fig. \ref{fig:alpres}). This large level of unresolved flux means that different
models of the system can satisfy the observations, with the main constraint being that
mid-IR emission must be detected only within a few AU of the star. Because this detection
could be the Wein side of the emission, it does not preclude the existence of material
just beyond a few AU. That is, the mid-IR emission does not indicate that no dust exists
where it was not detected, because it could be cool enough to evade detection.

The size of the 11$\mu$m images \citep[described in \S
\ref{ss:alp},][]{2010ApJ...723.1418M} compared to the reference PSF suggests a disk
radius of about 2.3AU. Given that this warm emission cannot be resolved by
\emph{Herschel}, there must either be (at least) two separate disk components (i.e. inner
and outer components), or a continuous disk that extends from a few to a few hundred AU,
whose structure appears different at different wavelengths due to the changing
temperature with radial distance.

As an example, the \emph{Herschel} data and SED can be modelled with an outer disk that
extends from 45-195AU with a decreasing face-on optical depth $\tau = 2.1 \times 10^{-5}
r^{-2.9}$, and an unresolved inner component at a temperature of 120K. The outer
component has a position angle of 345$^\circ$ and is edge-on, so consistent with being
aligned with the binary orbital plane. The interpretation of this model is that the
unresolved emission is distributed such that some dust is close enough to the star to
provide the mid-IR emission and structure, while most of it lies distant enough from the
star so that overall it appears to have a temperature of 120K. However, the distance
inferred for a blackbody at 120K around \alp~is 40AU, or 1\farcs8, which is relatively
large given the 6-7'' FWHM PACS beam at 70-100$\mu$m, and should be detectable in the
images.

We therefore use a more complex two power-law model, in which the disk extends
continuously from very near the star to several hundred AU. The surface density increases
from the inner edge to some break radius, beyond which it decreases to the outer
edge. This model has the advantage that it is extended near the binary, so may be able to
explain the mid-IR emission without invoking a separate warm component. Such a surface
density structure may be expected, because debris disk decay (and stirring in some
models) is an inside out process
\citep[e.g.][]{2004AJ....127..513K,2007ApJ...658..569W,2010MNRAS.405.1253K,2009MNRAS.399.1403M}.

Figure \ref{fig:alpres} shows the results of implementing this model, where the dust
extends from 1AU to 300AU, with the break radius at 50AU. The inner edge of 1AU
corresponds to the inner limit of stability for particles orbiting the \alp~binary
\citep[e.g.][]{1999AJ....117..621H}. The optical depth of the inner component is $\tau =
6.4 \times 10^{-8} r^{1.7}$. For the outer component $\tau = 3.3 \times 10^{-3}
r^{-1.6}$. The disk temperature has a blackbody dependence, with $T_{\rm disk} = 775
r^{-0.5}$K. As can be seen from the SED (Fig. \ref{fig:sed}), some modification of the
blackbody function is needed, with $\lambda_0=69\mu$m and $\beta=1.3$ for the inner
component and $\lambda_0=82\mu$m and $\beta=2.2$ for the outer component. The poor
resolution relative to the disk size means that these parameters are poorly constrained
with strong degeneracies. There is a factor of eight drop in surface density at the break
radius; attempts to keep the dust distribution continuous within the two power-law model
were unsuccessful because the outer component is fairly extended but with low surface
brightness. Joining the outer component smoothly to the inner component results in too
much flux just beyond the break radius. Such a drop is an expected feature of some disk
evolution models \citep[e.g.][]{2010MNRAS.405.1253K}, a point we return to in \S
\ref{s:disc}. The extent of the outer component is not very well constrained because the
surface brightness decreases strongly with radius. The disk model has a position angle of
345$^\circ$ so is consistent with the binary line of nodes. The disk is edge on, so again
consistent with being coplanar with the binary orbit. The images show significant
residuals for changes of 10-20$^\circ$ in both inclination and PA, giving an indication
of their uncertainties.

The two models considered above show similar residual structure when the model is
subtracted from the observations. The only sizeable ($>$3$\sigma$) residual related to
the star+disk is located on one side of the stellar position at 70$\mu$m. With the simple
models considered here, similar residuals remain for a range of PSFs (i.e. those shown in
Fig. \ref{fig:psf70}), even if the opening angle of the inner disk component is allowed
to vary (as might be expected if this component was misaligned with the outer component,
see \S \ref{s:dyn} below). Our inability to construct a model that accounts for these
residuals leads us to conclude that they are likely due to the beam variations discussed
in \S \ref{s:obs}. Alternatively, there may be additional structure unaccounted for by
our model, which would have a scale of a few arcseconds to leave such small
residuals. While the mid-IR images suggest that our coplanar configuration is
representative, these observations should be repeated, specifically at 18$\mu$m, as a
further test of disk coplanarity and structure near the star.

\begin{figure*}
  \begin{center}
    \hspace{-0.5cm} \includegraphics[width=0.5\textwidth]{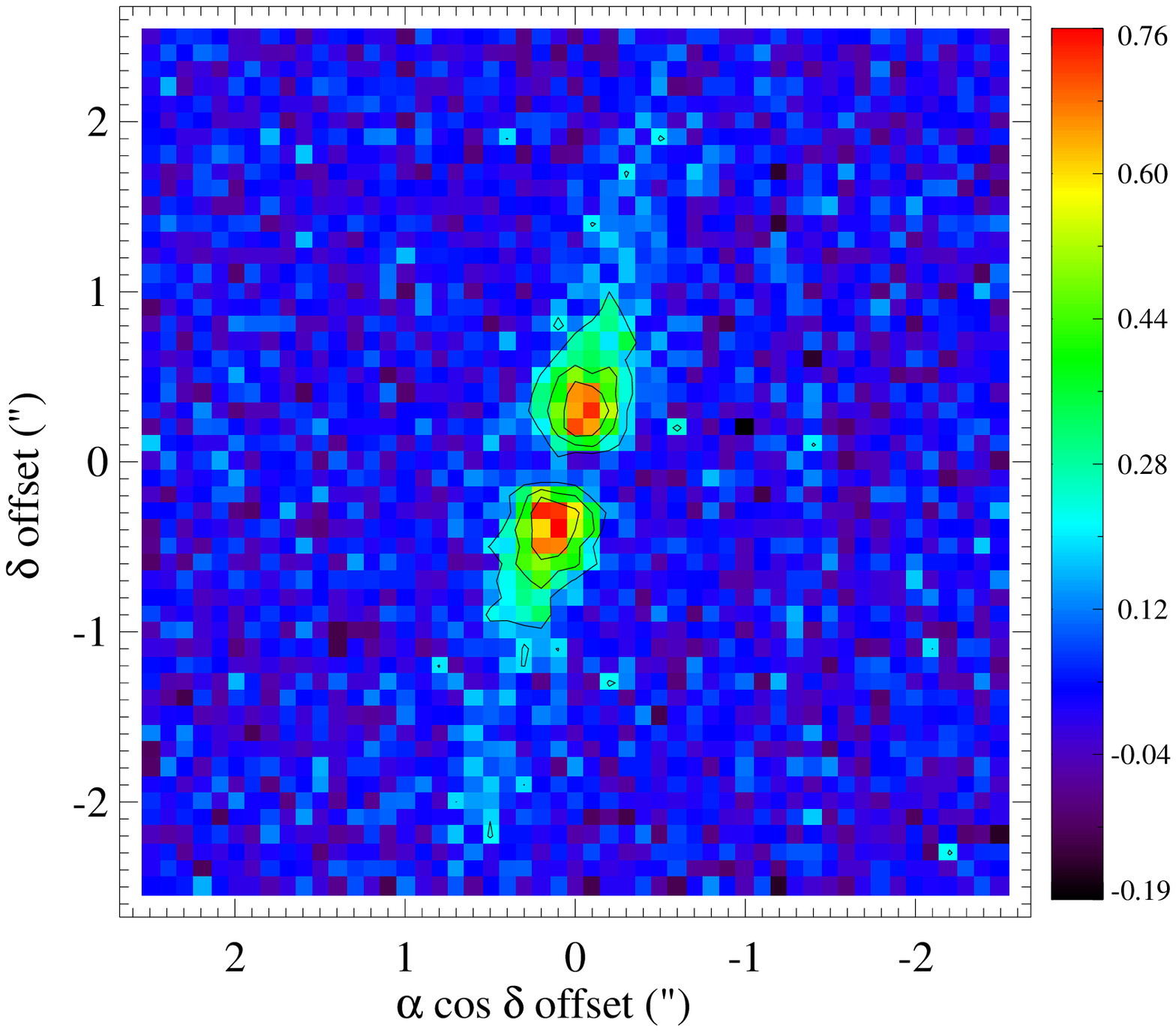}
    \hspace{0.1cm} \includegraphics[width=0.5\textwidth]{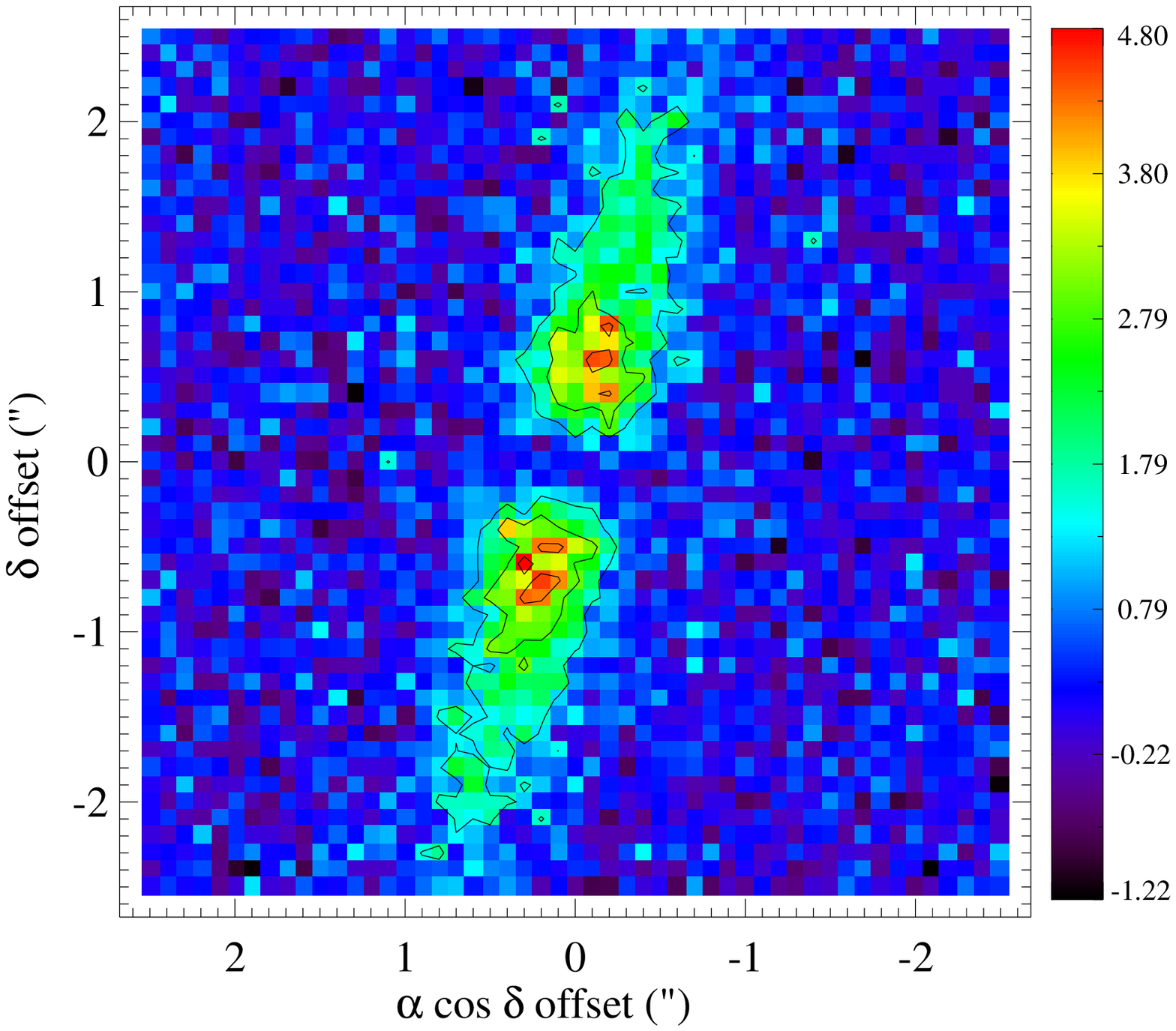}
    \caption{Simulated 11$\mu$m (left) and 18$\mu$m (right) peak-subtracted mid-IR
      observations of the \alp~model for comparison with \citet{2010ApJ...723.1418M}. The
      pixel scale is 0.1''/pixel, and the scales are in mJy/pixel. Contours are drawn at
      3, 6, and 9 times the background noise level. Our model has been multiplied by a
      factor of two at 11$\mu$m, and five at 18$\mu$m (see text).}\label{fig:midir}
  \end{center}
\end{figure*}

To compare the extended model with the resolved structure seen in the mid-IR by
\citet{2010ApJ...723.1418M}, Figure \ref{fig:midir} shows synthetic images at 11 and
18$\mu$m, where we have convolved the star+disk models with appropriately sized Gaussian
PSFs, added noise, and then subtracted the PSFs scaled to the image peak \citep[i.e. the
same method used by][to detect extension]{2010ApJ...723.1418M}. The background noise
level is calculated from the photometric uncertainties of 2 and 15mJy at 11 and 18$\mu$m
respectively, assuming aperture radii of 2\farcs0. In these images we have multiplied our
model by a factor of two at 11$\mu$m, and five at 18$\mu$m to make them appear similar to
the mid-IR observations, and show how the observed mid-IR structure could arise from an
extended disk. A factor of two at 11$\mu$m results in a disk flux of 65mJy, so is
reasonable given the uncertainty in the disk flux at this wavelength, which is limited by
the calibration of the photosphere and IRS spectrum. The factor five at 18$\mu$m results
in a disk flux of 1000mJy, which is less reasonable and would make the disk flux
inconsistent with the IRS and AKARI measurements. Our disk model is therefore fainter at
18$\mu$m than could be detected with the mid-IR imaging. That the 18$\mu$m observations
suffer from artefacts due to the observing procedure is the likely explanation, providing
further motivation for repeating these observations.

\subsubsection{\bet}

\begin{figure*}
  \begin{center}
    \hspace{-0.5cm} \includegraphics[width=0.33\textwidth]{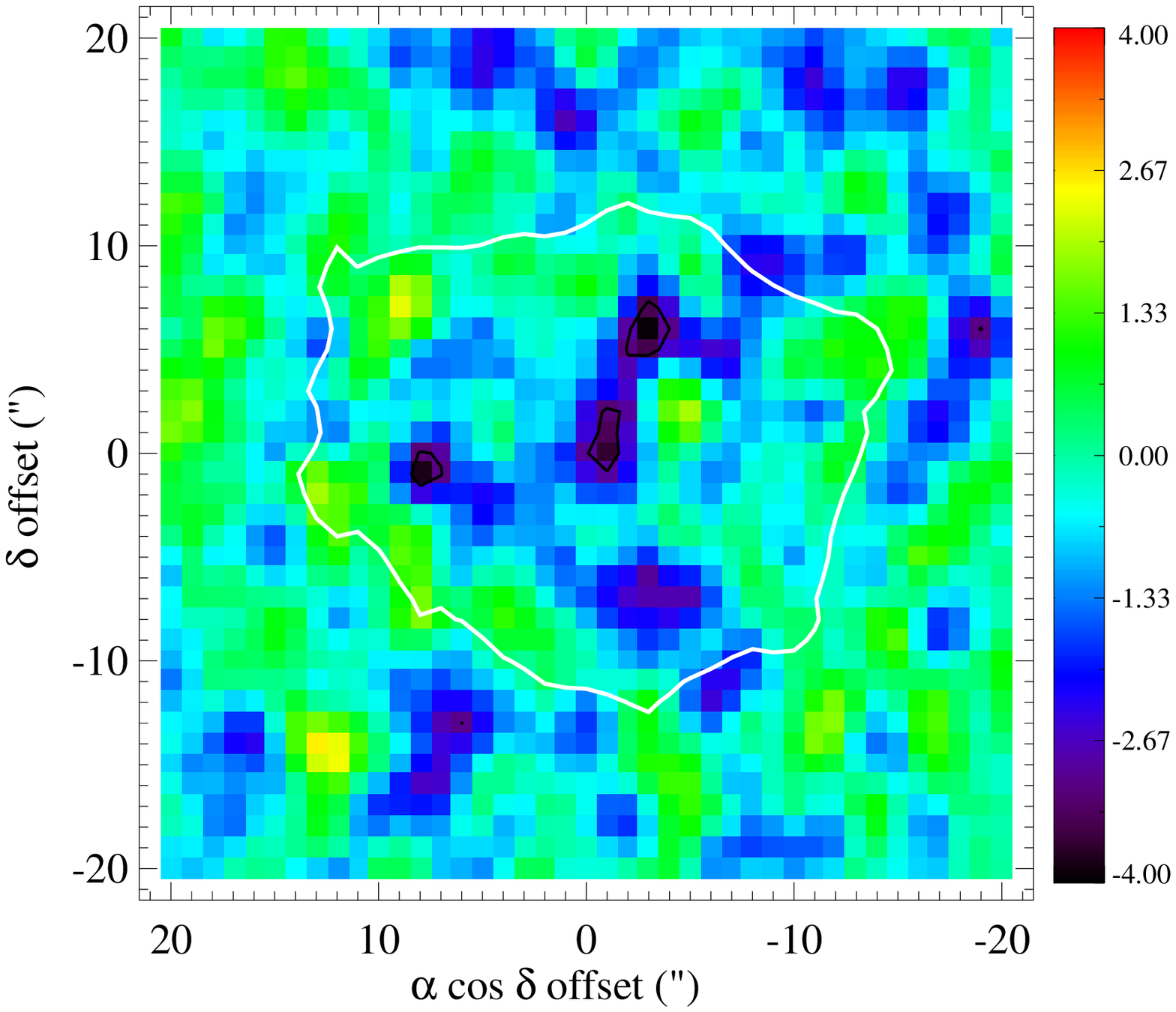}
    \hspace{0.1cm} \includegraphics[width=0.33\textwidth]{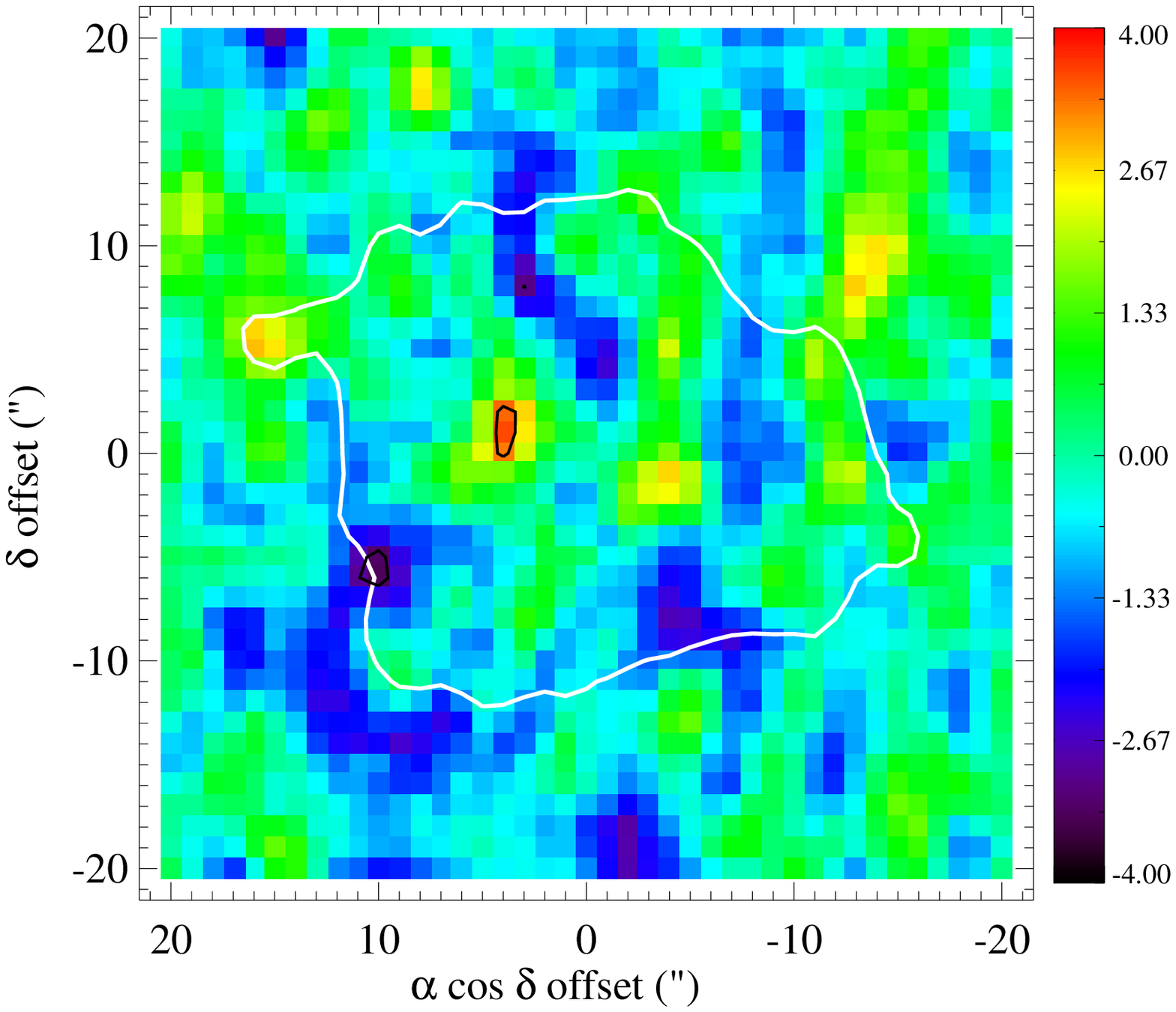}
    \hspace{0.1cm} \includegraphics[width=0.33\textwidth]{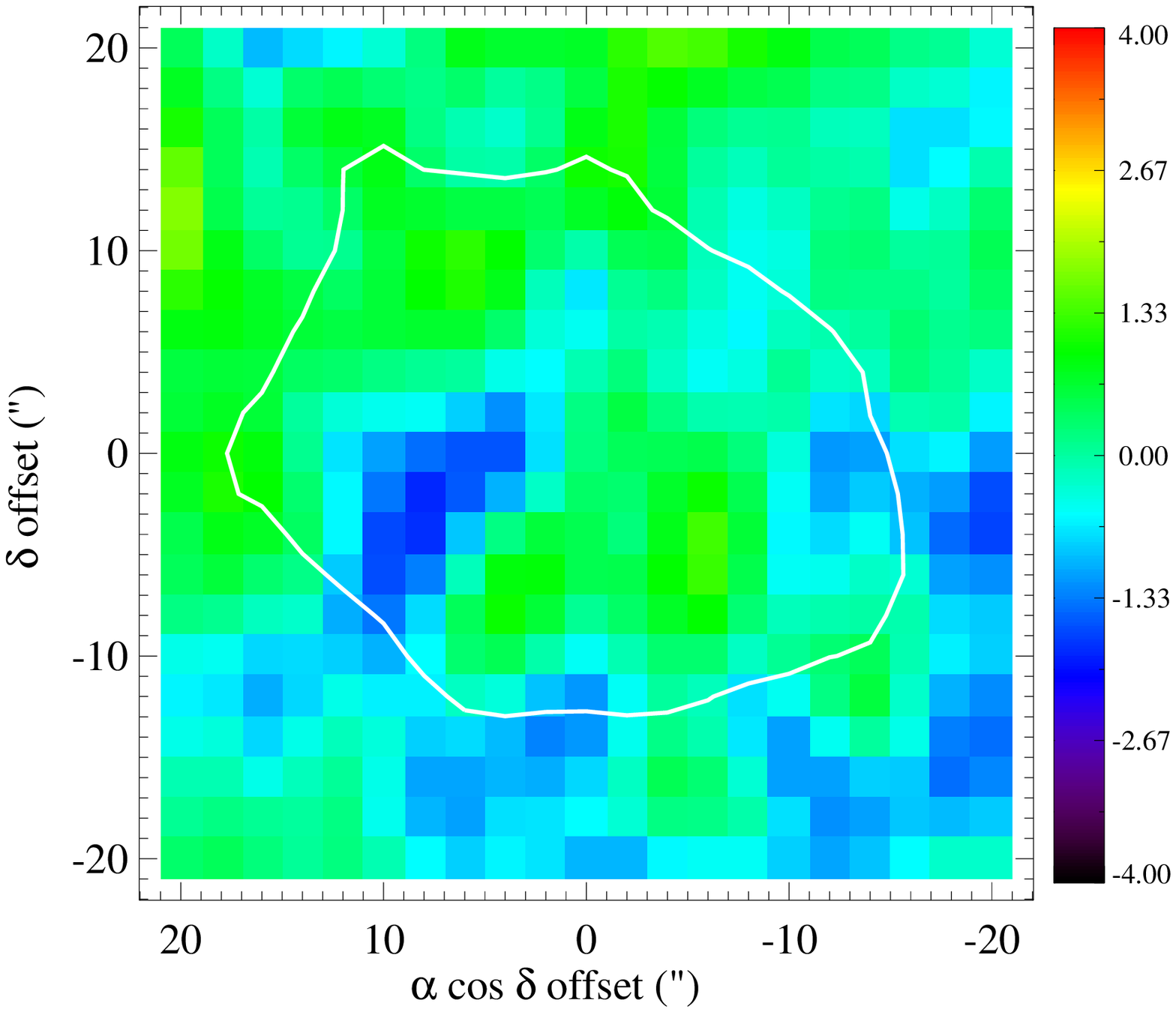}\\
    \hspace{-0.5cm} \includegraphics[width=0.33\textwidth]{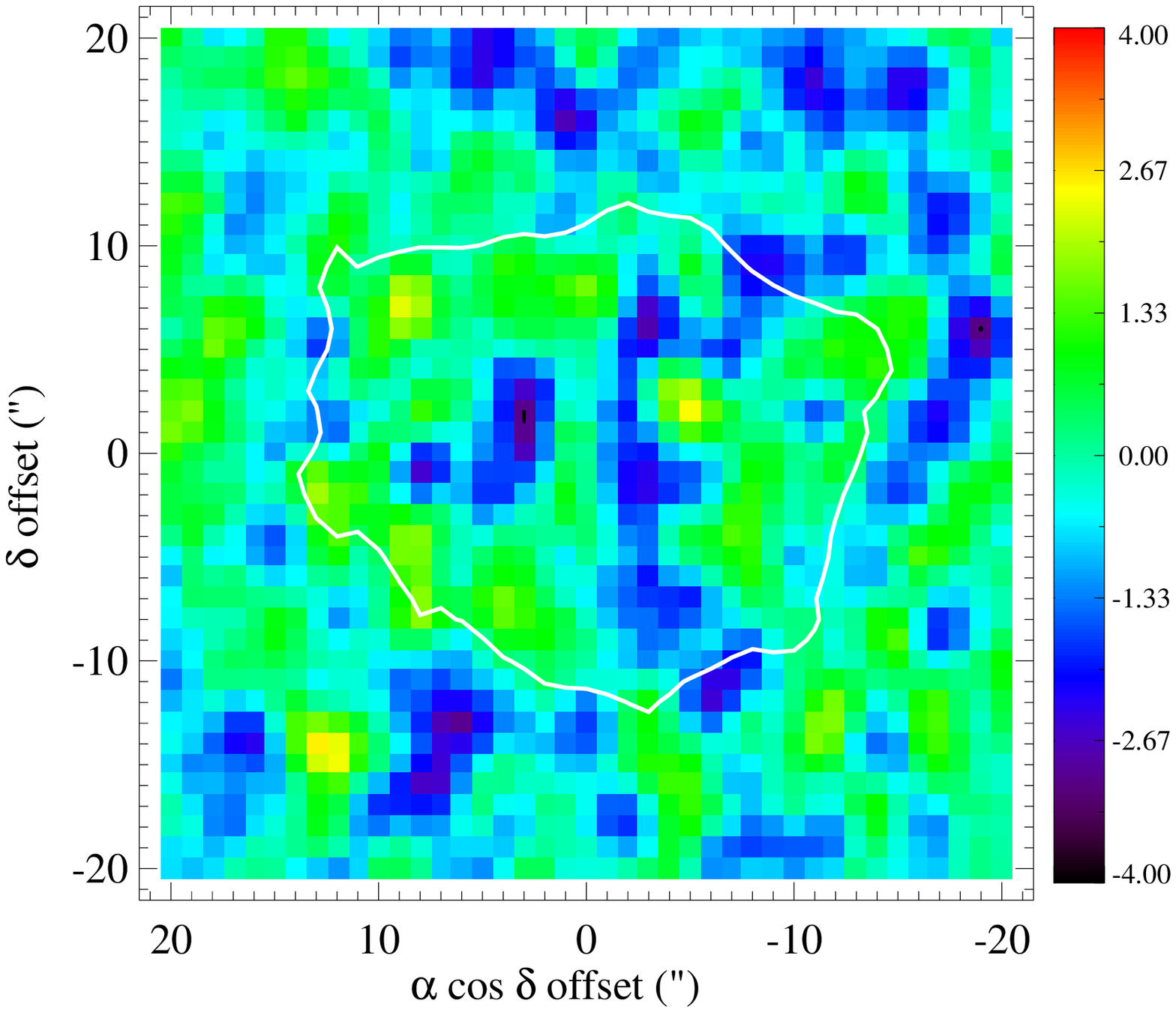}
    \hspace{0.1cm} \includegraphics[width=0.33\textwidth]{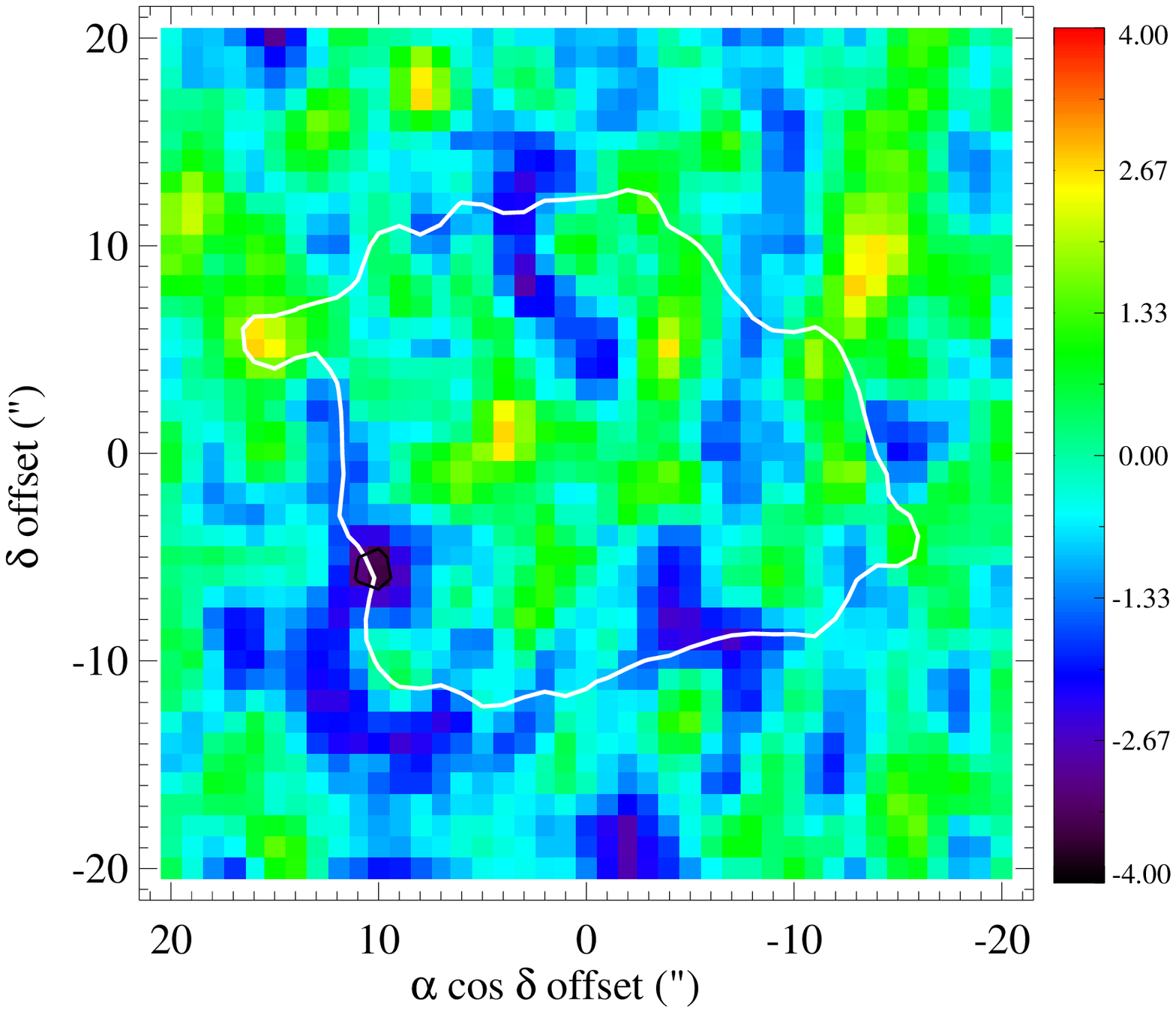}
    \hspace{0.1cm} \includegraphics[width=0.33\textwidth]{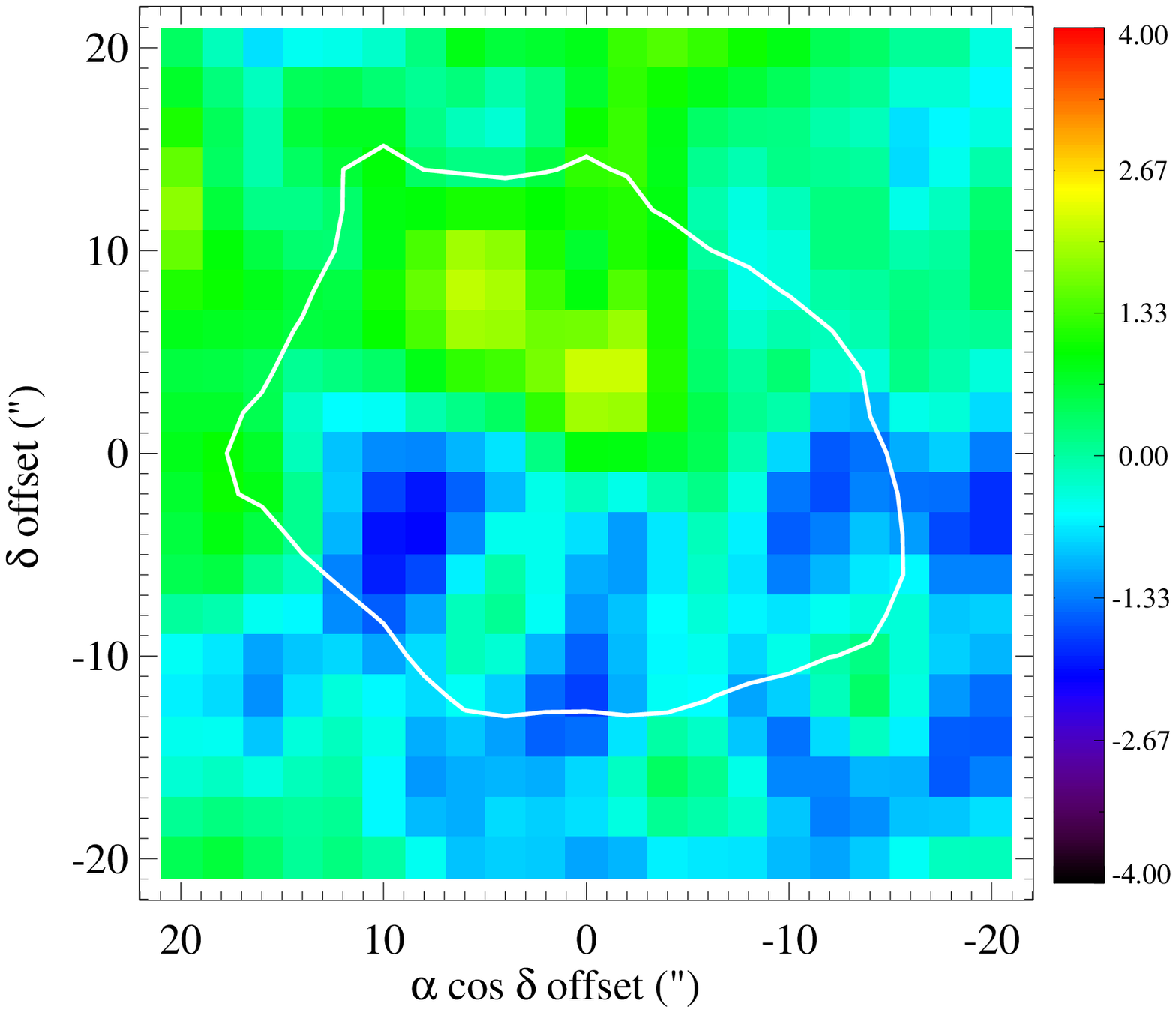}
    \caption{Residuals after the resolved model of \bet~is subtracted from the
      observations at 70$\mu$m (left) 100$\mu$m (middle) and 160$\mu$m (right). The top
      row shows results from the extended (50-400AU) model, and the bottom row shows
      results from the continuous (1-400AU) model. The colour scale is in units of the
      pixel to pixel uncertainty, and contours are shown at $\pm$3$\sigma$. The white
      contour shows the 3$\sigma$ contour from the observed images.}\label{fig:betres}
  \end{center}
\end{figure*}

The simplest \bet~model is shown in the top row of Figure \ref{fig:betres}. We found that
the images could not be modelled as a single narrow ring, so first allowed the inner and
outer disk edges to vary with a power-law surface density profile in between. This model
extends from 50-400AU with an optical depth of $\tau = 6.7 \times 10^{-5} r^{-0.9}$. The
temperature dependence is again that for a blackbody ($T_{\rm disk} = 819 r^{-0.5}$K),
with $\lambda_0=130\mu$m and $\beta=1.5$. The disk outer edge is not well constrained by
the PACS images, but is constrained to some degree by not being resolved with SPIRE at
250$\mu$m. The position angle and inclination are 67$^\circ$ and 46$^\circ$
respectively. Variation of the inclination by 10$^\circ$ leaves significant residuals
compared to the data, as does a similar change in the position angle. The position angle
of the binary plane is 65$^\circ$ so is easily consistent with that found for the
disk. The inclination of the binary plane is 40$^\circ$, so not significantly different
to that found for the disk. While the inclination is fairly well constrained the disk
opening angle is not, with values between 0-50$^\circ$ giving similarly good fits (though
lower disk opening angles are preferred). As we discuss in detail below in \S
\ref{s:dyn}, this ambiguity can be resolved by considering circumbinary particle
dynamics.

As with \alp, we also tried a continuous surface density profile by adding an inner
component to the first model. This continuous model extends from the innermost stable
orbit at about 4AU \citet{1999AJ....117..621H} out to 500AU with the surface density
peaking at 100AU and is equally successful at reproducing the \emph{Herschel} images
(bottom row of Fig \ref{fig:betres}). In this model we restricted the inner component to
join smoothly to the outer component; without this restriction the inner component can be
arbitrarily small and the model no different to the extended one above. The inner
component has an optical depth of $\tau = 2 \times 10^{-8} r^{1.7}$, and the outer
component decreases with an index of -1.7. The temperature is for a blackbody as above,
with $\lambda_0=85\mu$m and $\beta=1$ for the inner component and $\lambda_0=137\mu$m and
$\beta=1.5$ for the outer component. As with \alp, these parameters are poorly
constrained and highly degenerate.

While these two models show that the radial and vertical disk structure is poorly
constrained, both suggest that the disk optical depth decreases beyond a maximum that
lies around 60-100AU. Interior regions may be depleted as in the continuous model, or
completely empty. Despite these uncertainties, our main conclusion that the disk is
consistent with being aligned with the binary orbital plane is robust.

\section{Dynamics}\label{s:dyn}

The main result of the resolved modelling is that the disks are both consistent with
being aligned with the orbital planes of the host binaries. Whether this alignment simply
reflects the initial debris disk+binary configuration at the end of the protoplanetary
disk phase, or subsequent evolution requires some study of the expected dynamics.

Disk particles exterior to a pair of orbiting bodies (in our case a binary star system)
are subject to secular perturbations that modify the particle orbits on long
timescales. How the particle orbits are modified depends on the properties of the binary,
particularly the mass ratio and binary eccentricity, but also depends on the particle
orbits themselves \citep{2009MNRAS.394.1721V,2010MNRAS.401.1189F,2011arXiv1108.4144D}. If
the stars in the binary have similar masses, the effect on particle eccentricities is
smaller than for larger mass ratios \citep{2004ApJ...609.1065M,2012MNRAS.421.2264K}. In
contrast, the changes in particle inclinations and nodes can be significant. If the angle
between the particle and binary orbital planes is small (i.e. particles have low
inclinations with respect to the binary plane, we quantify ``small'' and ``large''
inclinations below), particle inclinations oscillate about the binary plane. Therefore,
while a disk of such particles will have an opening angle double the initial
misalignment, it will also appear to be aligned with the binary plane. The best example
is the warp in the $\beta$ Pictoris disk, which was proposed to be due to perturbations
by a planet misaligned with the disk plane
\citep{1997MNRAS.292..896M,2001A&A...370..447A} that was subsequently discovered
\citep{2009A&A...493L..21L,2010Sci...329...57L}. For large differences between the
particle and binary planes (i.e. high inclinations with respect to the binary plane),
particle inclinations are coupled to the evolution of their line of nodes, and oscillate
around a polar orbit. In the context of a debris disk these particles evolve into various
structures depending on the degree of initial misalignment
\citep{2012MNRAS.421.2264K}. Within the high-inclination family of orbits, only disks
that are initially misaligned by about 90$^\circ$ and have their line of nodes
perpendicular to the binary pericenter will not be strongly perturbed and continue to
appear disk-like.

The dividing line between ``large'' and ``small'' relative inclinations, and thus the
different families of disk structures, was quantified by \citet{2010MNRAS.401.1189F}. For
$\alpha$ CrB with $e=0.37$ the critical angle is 48$^\circ$, for \bet~with $e=0.433$ the
critical inclination is 43$^\circ$ (for an ascending node of $\pm$90$^\circ$). Thus, if
the disk inclination relative to the binary plane was initially larger than 48 and
43$^\circ$ respectively for these systems, it could not become aligned with the binary
plane.

\begin{figure*}
  \begin{center}
    \hspace{-0.5cm} \includegraphics[width=0.5\textwidth]{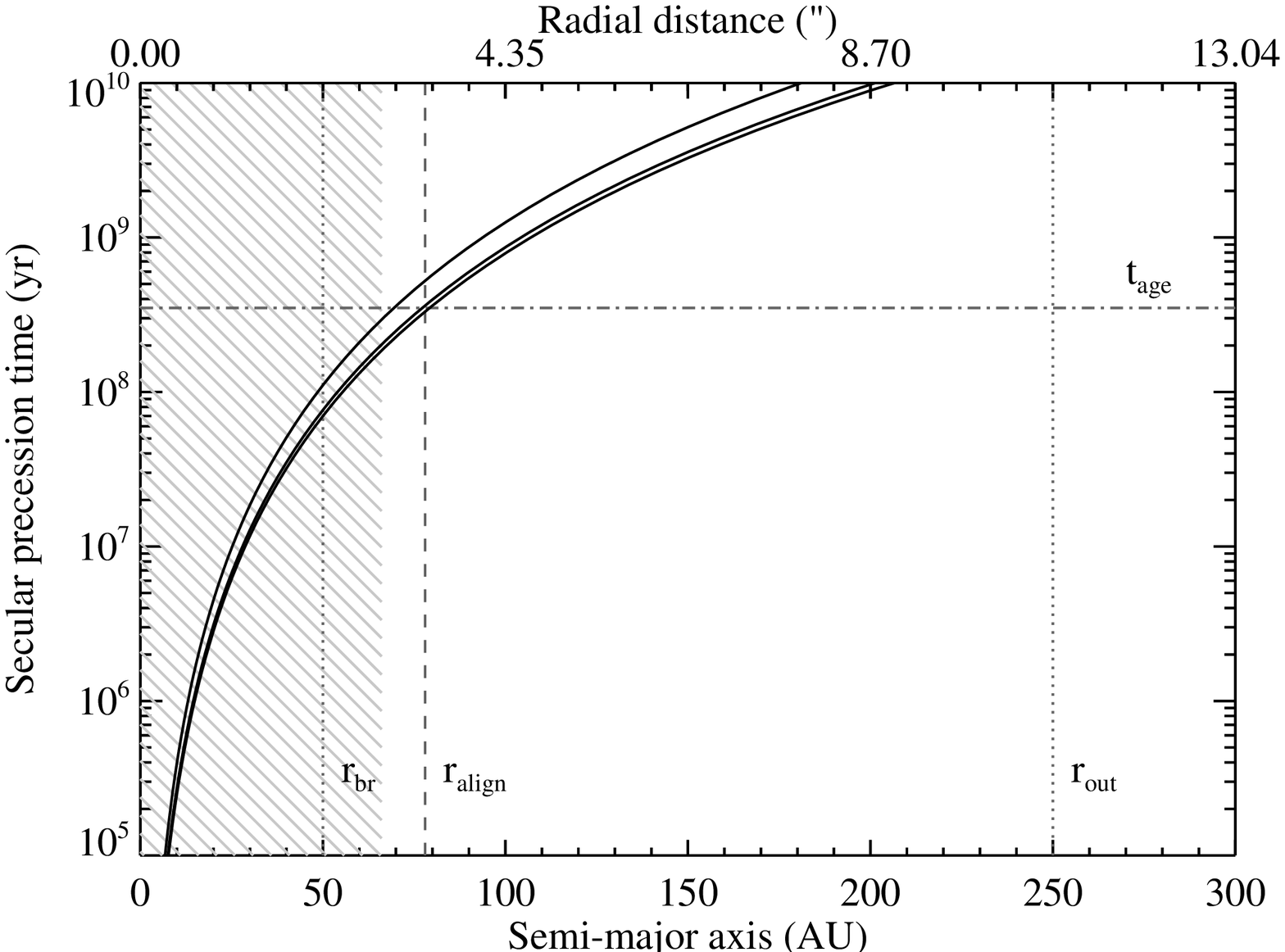}
    \hspace{0.1cm} \includegraphics[width=0.5\textwidth]{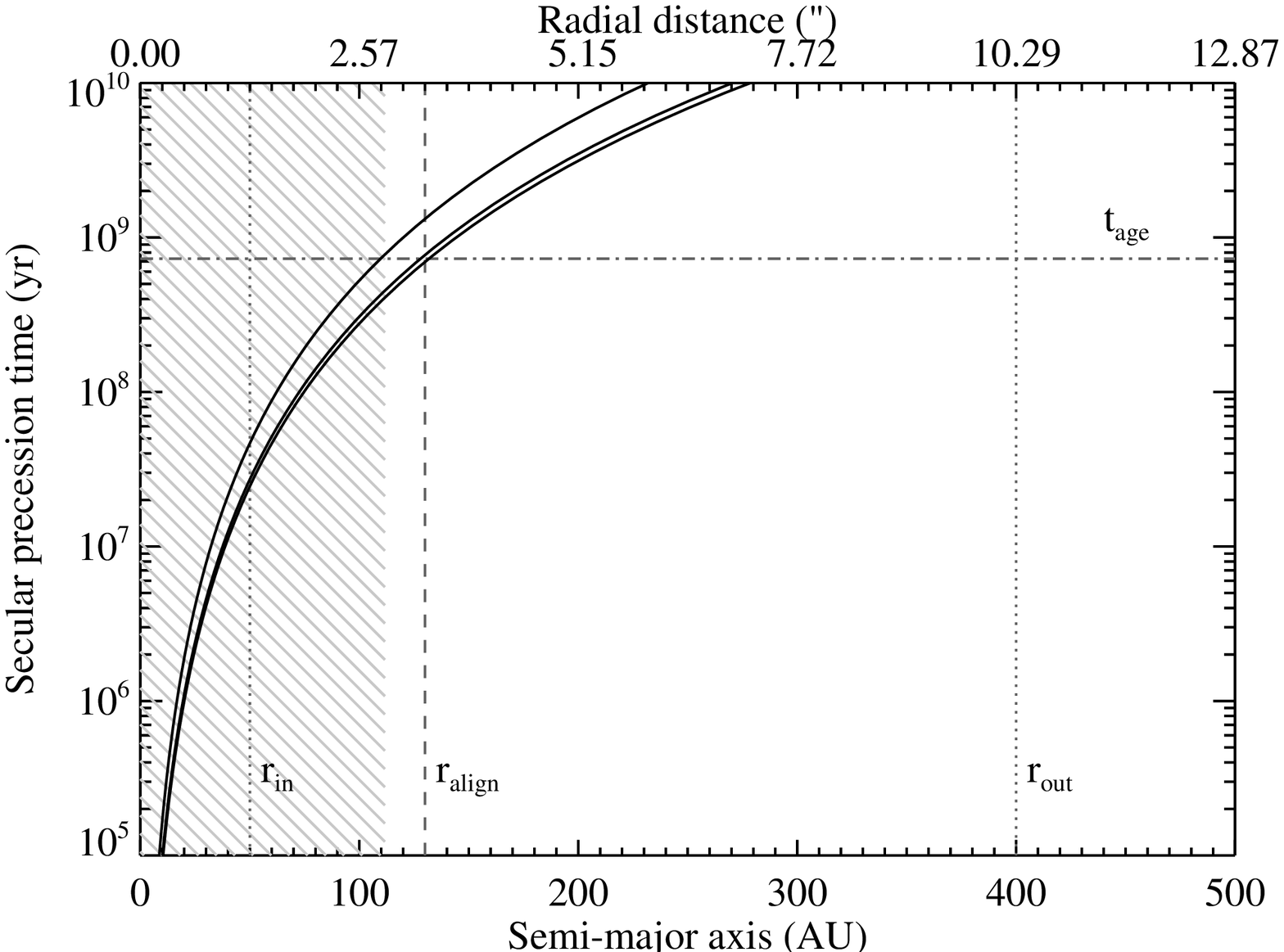}
    \caption{Secular precession times for \alp~and \bet~(solid lines, for misalignment
      angles of 0, 20, and 40$^\circ$ from bottom to top). The dot dashed line shows the
      estimated stellar age, and the dotted lines disk radii (the break and outer radii
      for the continuous \alp~model, and the inner and outer radii for the extended
      \bet~model). The alignment radius is where the secular precession time equals the
      stellar age for low misalignment angles. The hatched region lies inside the PACS
      beam half-width half-maximum at 70$\mu$m, so structure in this region is poorly
      constrained by PACS observations.}\label{fig:sec}
  \end{center}
\end{figure*}

While particles with initial inclinations lower than the critical angle can become
aligned, the finite secular precession time means that they will only be aligned if they
have been perturbed over a sufficiently long period. The timescale for alignment is given
by \citet{2010MNRAS.401.1189F}, and is shown for an arbitrarily small initial disk-binary
plane inclination, and for initial misalignments of 20 and 40$^\circ$ for \alp~and
\bet~in Figure \ref{fig:sec}. Particles with semi-major axes that lie to the left of
where the curves intersect a given system age (i.e. less than $r_{\rm align}$) have
completed at least one cycle of secular evolution. The hatched region indicates the
region where PACS cannot resolve the disk at 70$\mu$m, so shows where the disk structure
is poorly constrained by \emph{Herschel} observations.

The \alp~system is about 350Myr old, so the inclinations of particles within about 80AU
will be symmetric about the binary orbital plane. Beyond this distance however, particles
have not yet had time to undergo a complete cycle of secular precession and will retain
their original orbital inclinations. The transition between these regions is not sharp
due to the finite secular precession time \citep[see Fig. 1
of][]{2001A&A...370..447A}. Figure \ref{fig:sec} also shows that the alignment distance
is about the same as the inner extent of the PACS resolution at 70$\mu$m. Thus, the disk
structure that is constrained by the PACS images lies beyond $r_{\rm align}$, and the
alignment with the binary plane cannot arise due to secular perturbations. This
conclusion is strengthened by the finding that the orientation as seen by \emph{Herschel}
is consistent with the inner disk as seen in the mid-IR.

In the \bet~disk, particles beyond $r_{\rm align} \approx 140$AU are too distant to be
significantly affected by secular perturbations from the binary, assuming an age of
730Myr. Because we find that the disk extends well beyond this distance, the alignment of
the binary orbital and disk planes are again most likely primordial. Though we could not
constrain the disk opening angle well, the assumption that it is small is reasonable
based on i) observations that edge-on debris disks typically have small opening angles
\citep[e.g.][]{2005AJ....129.1008K,2006AJ....131.3109G}, and ii) that secular
perturbations could not have increased the scale height at the distances resolved by the
\emph{Herschel} observations.

\section{Discussion}\label{s:disc}

By considering the resolved models and expected dynamics, we have shown that the
circumbinary debris disks in both the \alp~and \bet~systems probably formed coplanar with
their parent binaries. A corollary is that because the disks were primordially aligned,
there should be little vertical structure inside the alignment radius. Both disks were
successfully modelled as disks with a single plane of symmetry, but are too poorly
resolved with \emph{Herschel} to strongly verify this statement. However, for \alp~the
low-level contours from the 11 and 18$\mu$m mid-IR imaging suggest that the inner disk
has a similar position angle to the binary and outer disk. Only one of the 18$\mu$m
images shows the same position angle, though the fact that the other is actually narrower
than the PSF in the same direction makes this extension questionable, as discussed by
\citet{2010ApJ...723.1418M}.

While there appears to be no need or evidence for vertical disk structure induced by
perturbations to disk particles' inclinations, an indirect signature could exist from
inclination and eccentricity variations imposed on disk particles. These variations
``stir'' the disk, where we are defining ``stirring'' to be any mechanism that increases
relative velocities between particles sufficiently that collisions become
catastrophically destructive, thus initiating a collisional cascade. In stirred regions
mass at the small end of the collisional cascade is removed by radiation forces, thus
depleting the disk. In unstirred regions the mass remains constant. The size distribution
in stirred regions has many more small grains for a given mass in large objects, so is
more easily detected due to the larger emitting surface area.\footnote{Once catastrophic
  collisions are occurring, the level of stirring is also important. An increase in
  stirring does not have a significant effect on the collision rate between particles of
  similar sizes because higher velocities are accompanied by a lower volume density of
  particles (e.g. the scale height and/or radial disk extent increases with the
  velocities). However, particles are typically destroyed in a collision with a much
  smaller particle, whose size is set by the collision velocity. Thus, higher levels of
  stirring mean that smaller particles are capable of destroying objects. In standard
  collisional size distributions \citep[e.g.][]{1969JGR....74.2531D} the number of
  particles, and therefore the number of potential destructive impactors, increases
  strongly with decreasing size, so the disk depletion rate due to collisions increases
  with increased stirring.}

Though the structure of the inner disk is not well constrained, we have shown that a
plausible model that satisfies both the \emph{Herschel} and mid-IR data for \alp~has a
continuous optical depth profile that peaks around 50AU. Because the typical expectation
from Solar System and protoplanetary disk studies is for the surface density to decrease
with distance from the star \citep[e.g.][]{1977Ap&SS..51..153W,2009ApJ...700.1502A}, this
profile can be interpreted as a depletion of material inside 50AU. Such a depletion is
expected in standard models of collisional evolution, where the disk decay rate is faster
at smaller semi-major axes \citep[e.g.][]{2003ApJ...598..626D,2007ApJ...658..569W}. This
50AU turn-over distance is similar to $r_{\rm align}$ shown in Figure \ref{fig:sec}, so
it is therefore plausible that the depletion inside here is due to increased velocities
imposed by binary perturbations.

Considering this binary stirring picture in more detail, vertical (inclination) stirring
is probably more important than radial (eccentricity) stirring in circumbinary disks,
particularly when the disk lies at large distances relative to the binary separation. As
the binary mass ratio decreases (i.e. the stars' masses become more similar), the
eccentricities imposed on an exterior particle decrease. For example, using the
expression derived by \citet{2004ApJ...609.1065M}, the ``forced'' eccentricity of
particles at 50AU around \alp~is about 0.001, at least an order of magnitude lower than
the eccentricities thought to exist in observed debris disks
\citep[e.g.][]{2004AJ....127..513K,2008ApJ...687..608K}. In contrast, for inclination
stirring to be effective the initial disk plane would only need to be a few degrees
different to the binary plane (though some non-zero eccentricity is needed to ensure
crossing orbits). For example, with an initial misalignment of 1$^\circ$, collision
velocities are stirred to about 0.02 times the Keplerian velocity (150m s$^{-1}$ at
50AU), so objects with dispersal thresholds less than roughly $10^4$J kg$^{-1}$
($\lesssim$10km) will be disrupted and dispersed in collisions with similar sized
objects. Therefore, the disk decays from the inside-out as secular perturbations increase
the relative velocities in particle collisions by aligning the disk, and do so on a
shorter timescale closer to the central star(s). Assuming that the initial misalignment
was large enough, collisions will be destructive inside some radius that will be close to
$r_{\rm align}$ \citep[but not exactly at $r_{\rm
  align}$,][]{2009MNRAS.399.1403M}. Outside this radius, collisions are unaffected by the
presence of the binary and therefore not destructive, providing a possible explanation
for the drop in optical depth in our model beyond 50AU.


While the \alp~disk is consistent with being stirred by the binary, we also consider two
alternative stirring mechanisms. The first is ``pre-stirring'' in which objects are
assumed to have been stirred at some very early time by an unspecified mechanism that
does not necessarily still operate (e.g. a stellar flyby or the result of gas disk
dispersal). A potential issue with such a scenario is that collisional damping may reduce
the velocities sufficiently that collisions are not catastrophic for a significant
fraction of the stellar main-sequence lifetime
\citep[e.g.][]{2002AJ....123.1757K,2004ARA&A..42..549G}. Whether collisional damping is
important depends on the relative sizes of the objects and their destructive impactors,
which in turn depends on the relative velocities. If pre-stirred objects can reach
sufficiently high eccentricities ($\gtrsim$0.1, though the value depends on object
strength and stellocentric distance), then the disk can remain stirred for the stellar
lifetime \citep{2011ApJ...739...36S}. Such a disk is depleted radially from the
inside-out, again because the depletion rate is a strong function of radius. The type of
structure that is expected is therefore a radially increasing optical depth profile,
which turns over and decreases where collisions have yet to deplete the disk
significantly \citep[e.g.][]{2010MNRAS.405.1253K}. Using equation (6) from
\citet{2010MNRAS.405.1253K} we find for the best fit planetesimal properties from
\citep{2007ApJ...663..365W} that the \alp~disk would be depleted within 50AU if the disk
was stirred from an arbitrarily early time (e.g. pre-stirred) and was about 5-15 times
less massive than the solid component of the minimum-mass Solar nebula \citep[MMSN,][when
scaled linearly with the binary mass]{1977Ap&SS..51..153W}.\footnote{We are not
  necessarily suggesting that the solid mass in the primordial protoplanetary disk was
  similarly depleted relative to the MMSN, as the mass could for example have gone into
  building planets. However, given the large observed dispersion in protoplanetary disk
  masses in star forming regions, such a depletion is easily possible
  \citep[e.g.][]{2005ApJ...631.1134A}}

The second mechanism is ``self-stirring'', where random velocities are initially slow
enough that collisions result in accretion and growth. Once the largest objects reach
roughly Pluto-size, they increase the velocities of smaller planetesimals and initiate a
collisional cascade, and the production of visible levels of dust
\citep{2004AJ....127..513K}. Again, this process works in a radially inside-out fashion,
so a self-stirred disk looks similar to a pre-stirred one, with one key
difference. Because planetesimals have not been stirred outside where Pluto-sized objects
have formed, collisions do not result in a collisional cascade and the disk should show a
drop in optical depth beyond this distance \citep{2010MNRAS.405.1253K}. Using equation
(9) from \citet{2010MNRAS.405.1253K}, for Pluto-sized objects to stir the disk to only
50AU by 350Myr, the disk would have to be 2000 times less massive than a scaled MMSN
\citep[see also][from which the Pluto-formation and stirring times were
derived]{2008ApJS..179..451K}. A disk with such low mass would not be visible in a
self-stirring scenario \citep[e.g.][]{2008ApJS..179..451K}, which appears to disfavour
this scenario. However, this calculation assumed that the planetesimals started out with
1m-1km sizes. If the planetesimals were initially much larger, the time to form
Pluto-sized objects would also be longer \citep[e.g.][]{2010ApJS..188..242K}. This longer
formation time would mean that a more massive disk, which would be correspondingly
brighter and therefore more consistent with the observations, could form Pluto-sized
objects and stir the disk only as far as 50AU by 350Myr.

Therefore all stirring models appear consistent with the observed peak in surface density
at 50AU. However, the model derived in \S \ref{ss:img} required a drop in surface density
beyond 50AU, a feature expected in self-stirred and binary-stirred disks, which would
appear to disfavour a pre-stirred interpretation, but not discern between self and
binary-stirring. A caveat on this conclusion is that the disk structure is poorly
constrained by the low resolution of the observations, and the dust observed with
\emph{Herschel} may not trace the parent body locations, particularly in the outer
regions. For example, the peak at 50AU may simply represent the outer edge of a parent
body disk that is pre-stirred, and the (decreased) emission beyond 50AU could arise from
small grains originating at 50AU forced onto larger eccentric orbits by radiation
pressure \citep[e.g.][]{2003A&A...408..775T,2006A&A...455..509K}.

Regardless of the stirring mechanism, we can compare the disk structures with those
expected if they decayed from some arbitrarily large level. In this picture the face-on
geometrical optical depth (with the same assumptions used above) is
\begin{equation}\label{eq:tau}
  \tau = 8.9 \times 10^{-5} r^{7/3} M_\star^{-5/6} L_\star^{-0.5} t^{-1}
\end{equation}
where $M_\star$ is the stellar mass in Solar units and $t$ is the system age in Myr
\citep[][equation 8]{2010MNRAS.405.1253K}. The $r^{7/3}$ dependence gives the expected
radial profile of a disk whose planetesimal properties are the same everywhere. With
these assumptions, for \alp~the expected optical depth at 350Myr is $1.15 \times 10^{-8}$
at 1AU, which is somewhat smaller than the model value of $6.4 \times 10^{-8}$. Given
that there is considerable uncertainty in the planetesimal properties and the true dust
distribution, we do not consider this difference a cause for concern. The model does not
increase as strongly with radius as Equation (\ref{eq:tau}), so is below the expected
level outside a few AU anyway. That the disk model has a $r^{1.7}$ dependence rather than
$r^{7/3}$ could indicate that the planetesimal properties have a radial dependence, for
example that they become smaller or weaker at larger distances (the model parameters are
very uncertain however, so such a dependence is not required).

The \bet~data are consistent with a continuous optical depth profile that peaks around
100AU.  Given that the secular precession time depends strongly on semi-major axis, and
that the stellar age is uncertain, this distance is not significantly inside $r_{\rm
  align}$ and the disk is therefore plausibly binary stirred. The expected optical depth
at 730Myr is $3.8 \times 10^{-9}$ at 1AU, which is smaller than the model value of $2.0
\times 10^{-8}$. Again, the discrepancy is not particularly large given the model
assumptions and uncertainty. Using the same equations as above, the \bet~disk would be
depleted out to 100AU by 730Myr for a disk 1-5 times less massive than a scaled MMSN if
it were stirred from the outset. Stirring by Pluto-formation out to this distance only
requires a disk 400 times less massive than a scaled MMSN (but again could be more
massive if planetesimals are larger). Thus, the disk could be depleted out to 100AU by
collisional evolution, but the stirring mechanism is unclear. Unlike \alp, which has a
mid-IR detection, there is no evidence for warm emission in the \bet~system. Such a
detection could break the degeneracy in our models, which cannot tell whether the regions
inside 50-100AU are devoid of debris (e.g. due to dynamical clearing by planets), or
simply depleted by collisional evolution.

This ``standard'' picture of debris disk stirring and evolution is not the only
possibility. For example, catastrophic collisions may be caused due to crossing orbits in
a disk where self-gravity is important, with the additional possibility that such disks
may appear non-axisymmetric \citep{2012MNRAS.421.2368J}. It is also possible that some
observed debris disks are not stirred to catastrophic collision velocities at
all. \citet{2010MNRAS.401..867H} show that long-lived ``warm'' planetesimal disks could
exist, in which collisions are not typically disruptive. They suggest that a test for
such a scenario is that the disk spectrum should look like a blackbody, because the disk
particles required for such warm disks to survive are large enough that they act like
blackbodies (i.e. absorb and emit light efficiently). For the two systems considered here
the disk spectra appear to rule out such a scenario because they lie significantly below
pure blackbodies beyond wavelengths of about 100$\mu$m (Fig. \ref{fig:sed}), suggesting
that the particles are emitting inefficiently at long wavelengths and are predominantly
smaller than $\sim$1mm. However, we do not exclude the possibility that the disks are
``warm'', because the small grains that are observed could have been created in erosive
and bouncing collisions. Modelling the size distribution would make more quantitative
predictions to test this possibility.

Though both disks have extended or continuous dust distributions as observed by
\emph{Herschel}, the parent bodies may (or may not) be more localised. All the models we
considered find that the dust optical depth decreases with distance in the outer
regions. Qualitatively, this structure is expected when the parent bodies occupy a narrow
``birth'' ring and small grains are placed on eccentric and hyperbolic orbits
\citep[e.g.][]{2006ApJ...648..652S,2010ApJ...708.1728M}. However, it is also possible
that the observed extent reflects the underlying parent body distribution, as might be
argued for \alp, which has dust detected both near and far from the star. Alternatively,
the drop in optical depth beyond 50AU required by the continuous \alp~model may be a sign
that the parent planetesimals lie relatively close to the star and that the more distant
dust comprises small grains on eccentric and hyperbolic orbits.

Whether debris disks are typically rings or more extended structures is an open question
\citep[e.g.][]{2006ApJ...637L..57K}, in part because obtaining sufficiently high
resolution sub-mm observations, those most sensitive to larger grains, is
challenging. This ambiguity has only been overcome in a few nearby systems, where a
parent body ring has been resolved at sub-mm wavelengths and is seen to be narrower than
the radial extent of small grains
\citep{2004Sci...303.1990K,2012ApJ...749L..27W,2012ApJ...750L..21B,2012A&A...540A.125A}. With
the development of facilities such as the Atacama Large Millimeter Array (ALMA) and the
Northern Extended Millimeter Array (NOEMA), the detection and resolution of larger parent
body populations will become possible and will provide insight into the processes that
set where planetesimals form and reside.

Looking at the issue of alignment from a wider perspective, coplanarity is the expected
outcome for debris disks and planetary systems emerging from the protoplanetary disk
phase for small to medium binary separations
\citep[e.g.][]{2000MNRAS.317..773B}. However, only a few examples where the outcome can
be tested actually exist. Three systems with transiting circumbinary planets, in which
the stars are also eclipsing binaries, show that well aligned systems are a possible
outcome \citep{2011Sci...333.1602D,2012Natur.481..475W}. Further,
\citet{2012Natur.481..475W} find that the occurrence rate of aligned circumbinary planets
is probably consistent with the rate for circumstellar planets with similar properties,
suggesting that alignment is the typical outcome. However, they acknowledge significant
biases exist, and that further work is needed to understand the implications of these
discoveries for circumbinary planetary system alignment and frequency.

In the case of circumbinary debris disks, only four systems where the disk and binary
alignment can be tested exist; \alp, \bet, 99 Herculis \citep{2012MNRAS.421.2264K}, and
HD 98800\footnote{A weak accretion signature has recently been detected for $\sim$10Myr
  old HD 98800 \citep{2012ApJ...744..121Y}, suggesting that it lies somewhere between the
  protoplanetary and debris disk phases.}
\citep{2005ApJ...635..442B,2010ApJ...710..462A}. Of these, \alp~and \bet~are close
binaries with periods of several weeks and HD 98800 has a period of 314 days, while 99
Her has a semi-major axis of 16.5AU and a 56 year period. With only these systems we
cannot yet be sure of what trends will emerge. As hinted by the disk-binary alignment of
\alp, \bet, and HD 98800, and the misalignment for 99 Her, it may be that more widely
separated systems are more likely to be misaligned. It will also be interesting to test
whether stirring by secular perturbations from binaries is important for disk
evolution. This hypothesis could be tested by comparing disk sizes (and structure where
possible) with the radii at which secular perturbations can have reached within the
system lifetime for a larger sample.

\section{Summary}\label{s:sum}

We have presented resolved images of debris disks around the nearby close binary systems
\alp~and \bet. These systems are relatively unusual among binaries because their orbital
configurations are relatively well known, allowing a test for (mis)alignment between the
disk and binary orbital planes. In both cases we find that the disk and binary are most
likely aligned. Though secular perturbations can align systems over time, the bulk of the
resolved disk structure in these systems is too distant to be affected. Therefore, the
alignment is most likely primordial, suggesting that the binary + protoplanetary disk
system from which the debris disk emerged was also aligned. These initial conditions are
consistent with expectations of alignment in protoplanetary disk+binary systems where the
binary has a separation less than about 100AU \citep{2000MNRAS.317..773B}.

Secular perturbations from the binary could provide the stirring mechanism in
circumbinary disks, and both \alp~and \bet~are consistent with such a picture. However,
they are also consistent with other stirring mechanisms. While binary stirring
may happen, it cannot be the only mechanism because debris disks are observed with a
similar frequency in both single and multiple star systems \citep{2007ApJ...658.1289T}.

These two systems bring the number in which debris disk-binary alignment can be
tested to four. Three of these (\alp, \bet, and HD 98800) have orbital periods of several
weeks to a year and appear to be aligned, while 99 Her has a period of 50 years and is
strongly misaligned. It is too early to draw conclusions about typical outcomes in such
systems, but the results so far suggest that misalignment cannot be extremely rare, and
may be preferred in systems with wider binary separations.

\section*{Acknowledgments}

Our thanks to Jonti Horner for his comments on a draft of this article, and to the
referee for comments that improved the discussion and overall clarity. This research has
made use of the Washington Double Star Catalog maintained at the U.S. Naval
Observatory. This work was supported by the European Union through ERC grant number
279973.


\end{document}